\documentclass[11pt]{article}
\usepackage[T1]{fontenc}
\usepackage[utf8]{inputenc}
\usepackage{lmodern}
\usepackage{amsmath,amssymb,amsthm,mathtools}
\usepackage[margin=1in]{geometry}
\usepackage{microtype,booktabs,array,xcolor,enumitem}
\usepackage{tikz,flafter}
\usepackage[colorlinks=true,linkcolor=blue!40!black,citecolor=blue!40!black,urlcolor=blue!40!black]{hyperref}
\usepackage{fancyhdr}
\setlist{topsep=5pt,itemsep=3pt}
\newtheorem{theorem}{Theorem}[section]
\newtheorem{lemma}[theorem]{Lemma}
\newtheorem{proposition}[theorem]{Proposition}
\newtheorem{corollary}[theorem]{Corollary}
\theoremstyle{definition}

\theoremstyle{remark}

\newcommand{\E}{\mathbb E}

\newcommand{\supp}{\operatorname{supp}}

\hypersetup{pdftitle={Logarithmic Embeddings into Probabilistic Spanning Trees for Distances and Capacities},
 pdfauthor={Mohammad Hajiaghayi},
 pdfsubject={Low-stretch spanning trees, fundamental-cut capacities, and oblivious routing}}
\title{\Large Logarithmic Embeddings into Probabilistic Spanning Trees\\[3pt]
for Distances and Capacities}
\author{Mohammad Hajiaghayi\\
 University of Maryland, College Park\\
 \texttt{hajiagha@umd.edu}}
\date{November 2, 2026}
\newcounter{paperalgorithm}
\newenvironment{paperalgorithm}[2]{%
  \par\medskip\noindent\begin{minipage}{\linewidth}%
  \refstepcounter{paperalgorithm}%
  \hrule height .4pt\relax\smallskip
  \noindent\textbf{Algorithm \thepaperalgorithm.}\quad #1\label{#2}%
  \par\nobreak\smallskip\begingroup\small
  \setlength{\parindent}{0pt}\setlength{\parskip}{3pt}%
}{%
  \par\endgroup\vskip4pt\hrule height .4pt\relax
  \end{minipage}\par\medskip
}

\begin{document}
\maketitle

\begin{abstract}
We prove that every connected $n$-vertex graph admits a probabilistic
embedding into its spanning trees with expected distance stretch
$O(\log n)$.  This settles the logarithmic stretch conjecture of
Alon, Karp, Peleg, and West, whose spanning-tree game appeared at the
1991 DIMACS Workshop on On-Line Algorithms and in SIAM Journal on
Computing in 1995.  A weighted form of the theorem also yields a
distribution over spanning trees with capacities exactly equal to
their fundamental graph cuts.  Every tree dominates every graph cut,
every cut is approximated within $O(\log n)$ in expectation, and the
same capacity distribution defines an $O(\log n)$-competitive
fractional oblivious routing.  The distance and capacity distributions
may differ.  Diamond graphs give matching logarithmic lower bounds
for stretch, congestion, and cut distortion.

For rational input, both distributions are constructed deterministically
in polynomial time and use at most $m$ trees, where $m$ is the number
of graph edges.  We prove an $\Omega(m/\rho)$ support lower bound
for congestion $\rho$; in particular, $(1-o(1))m$ trees can be
necessary for near-unit congestion, even when the optimum tends to one.
Thus linear dependence on $m$ is unavoidable in this regime.

The weighted theorem combines a protected-geodesic petal decomposition
with a pathwise entropy bound for locally selected centers.  A local
averaging inequality derandomizes the radii, and an adaptive rational
packing procedure constructs the distributions.
\end{abstract}

\section{Introduction}
\label{sec:introduction}

Approximating a graph by a distribution of trees is a basic tool in
the design of algorithms.  A tree makes distances, cuts, and routing
particularly simple, while a distribution can avoid the distortion
forced on any one tree.  There are, however, two different notions
of a tree approximation.  An abstract tree metric may introduce
Steiner vertices and edges unrelated to the input graph.  A spanning
tree must use the original vertices and original edges, with their
given lengths.  The latter restriction is essential when a tree path
must be a path in the network itself.

Alon, Karp, Peleg, and West~\cite{AKPWConference,AKPW} formulated the
spanning-tree problem as a zero-sum game between a tree player and an
edge player.  Their extended abstract appeared at the 1991 DIMACS
Workshop on On-Line Algorithms, with proceedings published in 1992;
the full paper appeared in 1995.
They proved a bound of
$\exp(O(\sqrt{\log n\log\log n}))$ and conjectured an
$O(\log n)$ bound.  Their original payoff differs pointwise by at
most one from $d_{T,\ell}(u,v)/\ell_{uv}$, so the two game values
differ by at most one.  The modified game asks for a distribution
over spanning trees whose expected stretch is small on every original
edge.  An edge guarantee immediately extends to every pair of
vertices by summing along a shortest graph path.

Subsequent work brought the upper bound close to logarithmic.
Elkin, Emek, Spielman, and Teng~\cite{EEST} obtained average stretch
$O(\log^2 n\log\log n)$.
Abraham, Bartal, and Neiman~\cite{ABN} obtained stochastic stretch
$O(\log n\log\log n(\log\log\log n)^3)$.
The petal decomposition of Abraham and Neiman~\cite{AN} gives
average stretch $O(\log n\log\log n)$ and a construction with the
same logarithmic factors in its nearly linear running time.
Low-stretch trees also underlie preconditioners for fast solvers of
symmetric diagonally dominant linear systems~\cite{SpielmanTeng2014}.
The remaining gap is a geometric one: local partitions must be
chosen so that they can be reconnected using original edges without
losing the distance estimate.

We close this gap by proving the logarithmic bound for spanning
trees.  The theorem is weighted: the demand assigned to an edge is
independent of its length.  This is the form needed both for the
spanning-tree game and for a capacity decomposition.  Our
construction runs in polynomial time for arbitrary rational input.
Its running time is not nearly linear; the contribution is
the logarithmic approximation within the spanning-tree family and
an explicit polynomial construction.

\subsection{Results}

Let $G=(V,E)$ be a connected undirected graph, with $n=|V|\geq2$
and $m=|E|$.  Throughout the paper, a spanning tree has vertex set
$V$ and uses only edges of $G$.  For lengths $\ell_e>0$, write
$d_{G,\ell}$ and $d_{T,\ell}$ for graph and tree distances.
For capacities $c_e>0$ and a tree edge $f$, let $S_{T,f}$ be either
component of $T-f$.  Define
\begin{equation}
\label{eq:canonical-capacities-intro}
 c_T(f)=c_G(\delta_G(S_{T,f})),\qquad
 L_T(f)=
 \begin{cases}
 c_T(f),&f\in T,\\
 0,&f\notin T.
 \end{cases}
\end{equation}
Thus the tree topology is a subgraph, while its capacities are
determined by its fundamental cuts.  These are exactly the smallest
capacities on that topology that dominate every graph cut
(Proposition~\ref{prop:cap-minimal-cut-capacities}).

\begin{theorem}[Logarithmic spanning-tree decompositions]
\label{thm:main}
There is an absolute integer $C_0>0$ with the following property.
Put $\rho_n=C_0\lceil\log_2(2n)\rceil$ and fix
$0<\vartheta\leq1$.  Let $\alpha_n=(1+\vartheta)\rho_n$.
\begin{enumerate}[label=(\roman*)]
\item For every positive length vector $\ell$, there is a distribution
$q^{\rm dist}$ over spanning trees such that, for all $u,v\in V$,
\begin{align*}
 d_{T,\ell}(u,v)&\geq d_{G,\ell}(u,v)
       &&\text{for every supported }T,\\
 \E_{q^{\rm dist}}d_{T,\ell}(u,v)
       &\leq\alpha_n d_{G,\ell}(u,v)
       &&(u,v\in V).
\end{align*}
\item For every positive capacity vector $c$, there is a distribution
$q^{\rm cap}$ over spanning trees, equipped with
\eqref{eq:canonical-capacities-intro}, such that
\[
 \E_{q^{\rm cap}}L_T(f)\leq\alpha_n c_f
 \qquad(f\in E).
\]
Consequently, for every $A\subseteq V$,
\begin{align*}
 c_T(\delta_T(A))&\geq c_G(\delta_G(A))
       &&\text{for every supported }T,\\
 \E_{q^{\rm cap}}c_T(\delta_T(A))
       &\leq\alpha_n c_G(\delta_G(A))
       &&(A\subseteq V).
\end{align*}
Splitting every pair's demand among its tree paths with the fixed
probabilities $q_T^{\rm cap}$ gives an
$\alpha_n$-competitive fractional oblivious routing.
\end{enumerate}
Each distribution can be supported on at most $m$ trees.  For
rational input and rational $\vartheta$, it can be constructed
deterministically using at most
\[
 1+\left\lceil
 \frac{20m\lceil\log_2(2m)\rceil}{\vartheta^2}
 \right\rceil
\]
weighted-tree calls and polynomial additional work.  The complete
running time is polynomial in the binary input length and
$1/\vartheta$.
\end{theorem}

The capacity distribution is fixed before the cut or the demand
vector is chosen.  Its guarantee is
$\max_f\E[L_T(f)/c_f]$, and the routing uses the fractional
mixture of tree paths.  The theorem does not require each sampled
tree to have small congestion.  The distance and capacity
distributions may differ; this is necessary even when all graph
lengths and capacities are one
(Proposition~\ref{prop:cap-simultaneous-obstruction}).

The main ingredient is a deterministic weighted-tree theorem.
Given arbitrary demands $a_e\geq0$ and lengths $\ell_e>0$, it
constructs a spanning tree satisfying
\begin{equation}
\label{eq:weighted-target}
 \sum_{e=uv\in E}a_e d_{T,\ell}(u,v)
 \leq\rho_n\sum_{e\in E}a_e\ell_e.
\end{equation}
The geometric and entropy arguments proving this statement occupy
Sections~\ref{sec:audited-geometry}--\ref{sec:local-rule}.
We use symbolic constants there and give an explicit admissible
choice, including $C_0$, in Section~\ref{sec:choice-of-parameters}.
The distribution construction in
Section~\ref{sec:refined-rational-master} works with any weighted
response bound $\rho$, loses only a factor $1+\vartheta$, and
has a call bound independent of the numerical size of $\rho$.

\paragraph{Tightness on diamonds.}
The logarithmic lower bound for probabilistic tree metrics on
recursive diamonds is classical~\cite{GNRS}.
We give a direct spanning-tree proof in
Section~\ref{subsec:diamond-lower}.  For every spanning tree of the
height-$k$ unit diamond $D_k$, the sum of tree distances over the
$4^k$ original edges is at least $k4^k/9$.
The same inequality, by double counting, gives an
$\Omega(\log n)$ lower bound for maximum expected
fundamental-cut congestion.  Summing singleton tree cuts gives
the lower bound for expected cut distortion as well.
Thus all three logarithmic bounds in Theorem~\ref{thm:main} are
asymptotically tight on the same series-parallel family.

\paragraph{The size of an explicit decomposition.}
Charikar, Chekuri, Goel, Guha, and Plotkin~\cite{CCGGP} showed that
$O(n\log n)$ abstract tree metrics suffice for their probabilistic
metric embedding.  This raises the corresponding support question
for spanning trees.  Our support bound $m$ cannot in general be
replaced by $O(n\log n)$ for fundamental-cut congestion, even when
the optimal congestion approaches one.
For an integer $k\geq2$, let $G_k$ consist of two $k$-vertex stars whose edges have
capacity $k^2$, together with all $k^2$ cross edges of capacity one.

\begin{theorem}[Support required by fundamental-cut congestion]
\label{thm:support-intro}
For $G_k$, every distribution with maximum expected congestion at
most $\rho$ has support at least $\lceil k^2/\rho\rceil$.
There is a distribution with congestion at most
$\beta_k=1+2(k-1)/k^2$.  Moreover, for every
\[
 \rho\geq\gamma_k=\frac{k^2+2k-2}{k^2-1},
\]
the minimum support size is exactly $\lceil k^2/\rho\rceil$.
On the same family, one fundamental-cut tree approximates every cut
within factor three; with unit metric lengths, one spanning tree
has distance distortion at most three.
\end{theorem}

The proof is in Section~\ref{sec:support-size-obstruction}.
Since $n=2k$ and $m=k^2+2k-2$, the theorem gives
$\Omega(m/\rho)$ support.  In particular, constant congestion
can require $\Theta(m)$ trees, and logarithmic congestion can
require $\Omega(n^2/\log n)$ trees.  The lower bound concerns
individual original-edge loads, the property used by the oblivious
routing guarantee.  It does not follow from cut approximation alone.
At the attainable bound $\rho=\beta_k=1+O(1/n)$, every distribution
needs at least $m-O(n)=(1-o(1))m$ trees
(Corollary~\ref{cor:support-near-full}).  Thus the dependence on $m$
is asymptotically sharp for congestion tending to one.

\subsection{Related work}
\label{subsec:related-work}

\paragraph{Probabilistic metric embeddings.}
Bartal~\cite{Bartal96,Bartal98} established probabilistic approximation
by hierarchically well-separated trees (HSTs), first with distortion
$O(\log^2 n)$ and then with distortion $O(\log n\log\log n)$.
Fakcharoenphol, Rao, and Talwar~\cite{FRT} obtained the optimal
$O(\log n)$ bound: every finite metric admits a distribution of
dominating tree metrics with this expected distortion for every pair.
Their construction also produces HSTs, whose edge lengths decrease
geometrically along root-to-leaf paths.
The distinction between deterministic domination and expected distortion
is central to the applications.  A fixed feasible solution has a cost
controlled in expectation by linearity, while domination supports
transferring a tree solution back to the original metric.
The allowed trees can have Steiner vertices and edges outside the input
graph.  Replacing such edges by graph paths can create cycles; extracting
a spanning tree from their union need not preserve the distance bound.
Consequently the metric theorem does not supply the original-edge
guarantee studied here.

\paragraph{Approximation and online algorithms.}
Metric labeling illustrates the role of HST structure.
Kleinberg and Tardos~\cite{KleinbergTardos} give a constant-factor
algorithm when the metric on the $k$ labels is an HST with a fixed
separation greater than two.  Combined with
FRT, this yields an $O(\log k)$ approximation for arbitrary label
metrics: assignment costs are unchanged, and the metric controls the
pairwise separation costs.
The related $0$-extension problem fixes the labels of designated
terminals.  Calinescu, Karloff, and Rabani~\cite{CKR} obtain an
$O(\log k)$ approximation, where $k$ counts terminals, through a
metric relaxation and randomized partitions.  Their partitioning method
is also an ingredient in the development of FRT~\cite{FRT}; this is a
connection between the techniques, rather than an attribution of the
$0$-extension theorem to FRT.

For network design, Awerbuch and Azar~\cite{AwerbuchAzar} use tree
embeddings for uniform buy-at-bulk, where cable types and their costs
per unit length are common to all edges.  Their constant-factor tree
algorithm and the FRT distortion bound give an $O(\log n)$ expected
approximation, including the multiple source--sink setting.
For group Steiner tree with $k$ groups, Garg, Konjevod, and
Ravi~\cite{GKR} obtain an $O(\log n\log k)$ approximation on trees;
combining it with FRT gives $O(\log^2 n\log k)$ on general graphs.
In online computation, Fiat and Mendel~\cite{FiatMendel} give an
$O(\log n\log\log n)$-competitive algorithm on HSTs.  Its classical
combination with FRT gives an $O(\log^2 n\log\log n)$ randomized
competitive ratio for metrical task systems against oblivious
adversaries~\cite{FRT}.
These examples explain why both the distortion and the structure of the
target trees matter: the reduction must preserve the particular
objective and feasibility conditions of its application.

\paragraph{Spanning trees and numerical algorithms.}
The original-edge problem has a separate history, beginning with the
tree--edge game of Alon, Karp, Peleg, and West~\cite{AKPW}, motivated
by the $k$-server problem on road networks.  Their reduction relates the
game value to a randomized competitive ratio against an oblivious
adversary.  The subsequent star decompositions of Elkin, Emek,
Spielman, and Teng~\cite{EEST}, the portal highways of Abraham,
Bartal, and Neiman~\cite{ABN}, and the petal decompositions of Abraham
and Neiman~\cite{AN} successively improved the distortion while
maintaining a spanning subgraph.  The latter construction achieves
$O(\log n\log\log n)$ average stretch in
$O(m\log n\log\log n)$ time.
For unweighted series-parallel graphs, Emek and Peleg~\cite{EmekPeleg}
already obtain a tight logarithmic distribution over spanning trees.
The diamond lower bound of Gupta, Newman, Rabinovich, and
Sinclair~\cite{GNRS} therefore holds even in a graph class where the
spanning-tree question was previously settled.

Low-stretch spanning trees also enter combinatorial preconditioning.
Spielman and Teng~\cite{SpielmanTeng2014} construct sparse subgraph
preconditioners by adding edges to low-stretch trees and use recursive
preconditioning to solve symmetric diagonally dominant systems in
nearly linear time.  This application requires efficient construction,
in addition to a stretch bound.  Our polynomial-time weighted oracle
establishes the logarithmic distortion guarantee; obtaining a comparable
running time remains a separate algorithmic question.

Recent work also studies faster constructions.
Kyng et al.~\cite{KyngEtAl2025} obtain efficient
logarithmic metric-tree embeddings and $\ell_1$-oblivious routing by
analyzing random-shift decompositions across scales.
Dorkofikis et al.~\cite{Dorkofikis2026} obtain spanning trees from
randomly perturbed shortest-path computations.  Their worst-case
construction has $O(\log^5 n)$ expected stretch and
$\widetilde O(m)$ running time for positive, polynomially bounded
integer lengths.  These results address computational efficiency and
alternative constructions alongside the distortion question.

\paragraph{Cut decompositions and minimum bisection.}
R\"acke~\cite{Racke} constructs a convex combination of capacitated
decomposition trees that supports $O(\log n)$-competitive fractional
oblivious routing in undirected networks.  The leaves correspond to
graph vertices, while tree edges are mapped to graph paths; these trees
are not required to be spanning subgraphs.
The same framework gives an $O(\log n)$ approximation for minimum
bisection with exact cardinality balance: when $n$ is even, both output
parts have exactly $n/2$ vertices, and the cut capacity is within the
stated factor of the minimum among such bisections.
R\"acke computes a minimum-cost leaf bisection in each supported tree
by dynamic programming and returns the best induced graph bisection
\cite[Section~3, Lemma~9]{Racke}.
The balance counts leaves, and hence original vertices; internal tree
nodes do not contribute to it.  The framework also yields logarithmic
approximations for sparsest cut and multicut.
Thus the logarithmic bisection approximation is an established
application.  Corollary~\ref{cor:cardinality-cuts} recovers this guarantee,
and its analogue for any prescribed side cardinality, using spanning
trees on the original vertex set with prescribed fundamental-cut
capacities.  The additional contribution concerns this representation
and its construction.

\paragraph{Family-preserving duality and explicit distributions.}
Andersen and Feige~\cite[Theorem~6]{AF} formulate the equivalence
between stretch and congestion for a fixed family of admissible
mappings.  Their discussion explicitly distinguishes the distributions
needed for the two guarantees.  For spanning trees, the mapping sends
each edge to its tree path, and the induced loads are precisely
fundamental-cut capacities.  They already use this family for a
near-logarithmic minimum-bisection approximation and explain how abstract
tree metrics recover R\"acke's logarithmic bound
\cite[Sections~4.5 and~5]{AF}.  Our weighted theorem supplies a
logarithmic response within the spanning-tree family itself.

Charikar, Chekuri, Goel, Guha, and Plotkin~\cite{CCGGP} give a
deterministic polynomial construction of a distribution on
$O(n\log n)$ metric trees with $O(\log n\log\log n)$ distortion.
Their work predates FRT and connects weighted response oracles to
explicit distributions through fractional packing.
Our rational master follows this framework and the multiplicative-weights
method~\cite{PST,AHK}; it gives a factor $1+\vartheta$ loss,
uniformly bounded query encodings, and exact compression to $m$ trees.
The support obstruction proved here concerns fundamental-cut congestion
on original-edge trees.  Its $\Omega(n^2/\rho)$ bound explains why the
$O(n\log n)$ support guarantee for metric approximation cannot be
transferred to that requirement, even at logarithmic congestion.

\paragraph{Organization.}
Section~\ref{sec:techniques} explains the proof and construction.
Section~\ref{sec:preliminaries} fixes conventions.
Sections~\ref{sec:audited-geometry}--\ref{sec:weighted-running-time}
give the weighted theorem and its running time.
Sections~\ref{sec:refined-rational-master} and
\ref{sec:capacity-extension} construct the distributions and prove
the capacity and routing statements.
Section~\ref{sec:lower-bounds} proves the lower bounds, and
Section~\ref{sec:choice-of-parameters} checks the constants.
\section{Our approach and techniques}
\label{sec:techniques}

Our starting point is a weighted formulation of the spanning-tree
problem.  Give each graph edge a positive length $\ell_e$ and an
arbitrary nonnegative demand $a_e$.  We construct one actual spanning
tree $T$ for which
\[
 \sum_{e=uv}a_e d_{T,\ell}(u,v)
 \leq O(\log n)\sum_ea_e\ell_e.
\]
The demands may depend on the lengths.  This freedom is what allows
the same theorem to yield both distance and capacity decompositions.
For distances, a weighted response controls a specified linear
combination of edge stretches.  For capacities, given positive edge
prices $p_e$, we use auxiliary lengths $p_e/c_e$ and demands $c_e$.
Double counting the resulting tree paths identifies the weighted
distance objective with the priced sum of fundamental-cut
congestions.  This is the family-preserving distance--capacity
equivalence of R\"acke and Andersen--Feige~\cite{Racke,AF}.
Our task is therefore to prove the weighted bound while keeping every
tree edge in the input graph.  The distance and capacity distributions
obtained from it may be different.

A demand routable in the graph with congestion $\gamma$ induces
load at most $\gamma c_T(f)$ on every tree edge, by the cut
inequality for its fundamental cut.  Averaging the corresponding
original-edge tree paths therefore gives the stated oblivious routing.

\subsection{The logarithmic weighted bound}

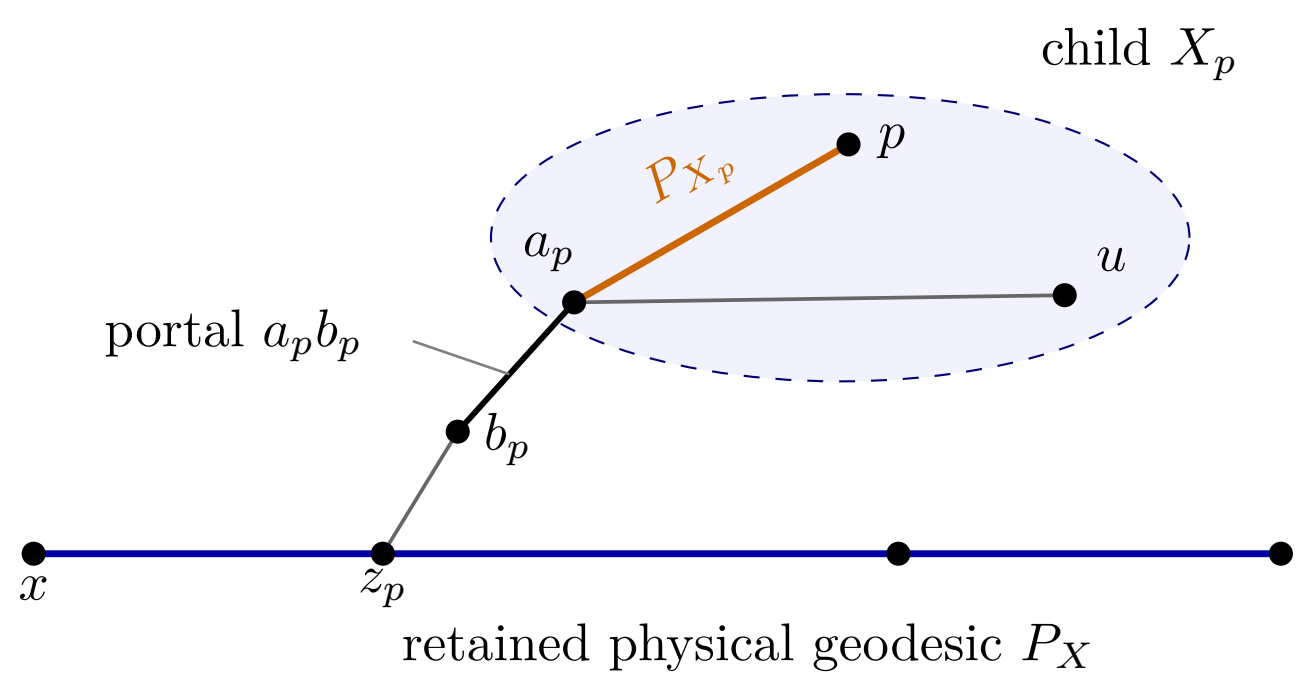
\begin{figure}[t]
\centering
\begin{tikzpicture}[x=1.10cm,y=0.95cm,
  vertex/.style={circle,fill=black,inner sep=1.6pt},
  every node/.style={font=\small}]
  \fill[blue!5] (4.85,2.20) ellipse (2.10 and 1.00);
  \draw[blue!45!black,dashed] (4.85,2.20) ellipse (2.10 and 1.00);
  \coordinate (x) at (0,0);
  \coordinate (z) at (2.10,0);
  \coordinate (q) at (5.20,0);
  \coordinate (y) at (7.50,0);
  \coordinate (b) at (2.55,0.85);
  \coordinate (a) at (3.25,1.75);
  \coordinate (p) at (4.90,2.85);
  \coordinate (u) at (6.20,1.80);
  \draw[blue!65!black,line width=1.3pt] (x)--(z)--(q)--(y);
  \draw[black!60,line width=0.7pt] (z)--(b);
  \draw[black,line width=1.1pt] (b)--(a);
  \draw[orange!80!black,line width=1.3pt] (a)--(p);
  \draw[black!60,line width=0.7pt] (a)--(u);
  \foreach \v in {x,z,q,y,b,a,p,u} {\node[vertex] at (\v) {};}
  \node at (0,-0.24) {$x$};
  \node at (2.10,-0.24) {$z_p$};
  \node at (2.85,0.80) {$b_p$};
  \node at (3.10,2.10) {$a_p$};
  \node at (5.16,2.87) {$p$};
  \node at (6.48,2.05) {$u$};
  \node[rotate=31,text=orange!80!black] at (3.95,2.65) {$P_{X_p}$};
  \node[anchor=east] at (2.10,1.52) {portal $a_pb_p$};
  \draw[black!50,line width=0.5pt] (2.28,1.48)--(2.86,1.25);
  \node at (4.30,-0.65) {retained physical geodesic $P_X$};
  \node at (6.65,3.48) {child $X_p$};
\end{tikzpicture}
\caption{A recursive step.  The child avoids the inherited protected
path and is joined through the original edge $a_pb_p$.  Its new
protected path runs from $a_p$ to the selected source $p$.
Only the protected paths, the relevant root-path segment, and one
branch of the child are shown; physical lengths are not to scale.}
\label{fig:techniques-protected-path}
\end{figure}

\paragraph{Protecting an entire shortest path.}
Recursive tree constructions must control the distances created when
their clusters are reconnected.  Petal decompositions achieve this by
coordinating cluster boundaries with paths leading toward a root
\cite{ABN,AN}.  We retain this geometric principle and change the
object carried by the recursion: each instance protects an entire
physical shortest path starting at its root.  Every child
avoids that path, which is retained verbatim in the final tree.
The child receives a new protected path, inherited from the root's
shortest-path tree through the chosen source; see
Figure~\ref{fig:techniques-protected-path}.

The construction uses temporary lengths that halve precisely the
protected-path edges.  A petal boundary is defined by a distance gauge
adapted to a shortest-path tree in these temporary lengths.  Removing
a petal leaves an ancestor-closed residual, so its surviving root
distances stay unchanged.  Each child has one designated original
portal edge leading to the later residual.  Its witness paths lie
entirely within the child and
contain no edge of the old protected path.  Consequently they have
their original physical lengths, and the new designated path is
physically shortest in the induced child.  These properties are
preserved for every admissible radius choice.

The resulting tree has constant distortion from every vertex of the
protected path to every vertex in the instance.  This gives the cost
scale needed in the analysis: an edge first separated in an instance
can be charged according to its endpoint's distance to that instance's
protected path.  Once this distance becomes comparable to the edge
length, the edge's entire remaining tree distance is paid immediately.
The geometry also permits every surviving off-path vertex to compete
as a source, which is essential for the local comparisons below.

\paragraph{Local scales and a carried threshold.}
For a vertex $v$ in an instance $X$, let $t_X(v)$ be its physical
distance to the protected path, measured at entry to the instance.
We fix an allowance $R_v=\varepsilon t_X(v)$, for a sufficiently small
absolute constant $\varepsilon$.  These allowances stay fixed while
the parent is carved.  Their Lipschitz property supplies meaningful
comparisons between nearby vertices even after residual distances
have increased.

We also fix a positive vertex measure once for the entire construction.
Write $b_e=a_e\ell_e$ and $B=\sum_e b_e$, and assume $B>0$;
the zero-demand case is immediate.  Assign each vertex the
baseline weights of its incident edges and an additional mass $B/n$.
Its total mass is $3B$, and its entropy is at most $3B\log n$.
This single entropy budget will pay for all charged edges throughout
the recursion.  In a current residual, the core of $v$ is a small
physical ball of radius $\delta R_v$; denote its mass by $m_v$ and
the fixed parent mass by $M$.

The personal rate of a vertex is proportional to
$K_v/R_v$, where $K_v=\lceil\log_2(2M/m_v)\rceil$.
A vertex is eligible when this rate is at most the current threshold
$\Lambda$.  Among eligible vertices we maximize $h(v)-R_v$, where
$h(v)$ is its current root distance.  The threshold doubles when no
candidate is eligible.  Each child starts with twice the threshold
used at its own source selection.  The priority rule ensures that an
eligible captured vertex has an allowance comparable to the source's
allowance and a child root distance at most $\varepsilon t_X(v)$.
Thus eligibility gives strong contraction; ineligibility gives a
mass inequality that will instead pay for the capture.

\paragraph{Summing over physical scales.}
The natural charge at a capture of $v$ is $\Lambda t_X(v)$.
Although $t_X(v)$ need not decrease at every capture, it always lies
within a constant factor of its running minimum.  The reason is that
current root heights decrease along a recursive history, and every
capture resets the next root height to at most a constant times the
present distance to the protected path.  This also controls any later
increase in that distance.

We group captures into dyadic bands of the running minimum.  Within
one band, the physical scales are comparable while the carried
thresholds grow geometrically.  The total charge in that band is
therefore bounded by a constant times its last charge.  For these
representatives, place an inner and an outer physical ball around the
vertex.  Their radii are fixed constant multiples of the running
minimum.  After a constant number of occupied bands, the later outer
ball is contained in the earlier inner ball.  Residuals are nested,
and intrinsic distances cannot decrease under restriction.  The
logarithms of these local mass ratios consequently telescope,
independently of the number of recursive levels or the metric aspect
ratio.

\paragraph{Far captures and inherited rates.}
Call a capture far when the source allowance is at most $1/\chi$
times the captured vertex's allowance, for a fixed constant $\chi\geq4$.
Such a vertex must be ineligible:
the comparison for eligible captures rules out this discrepancy.
Using $K_v$ within constant factors of $\log(2M/m_v)$, the two
eligibility comparisons bound the charge by a difference of logarithms.
Their common parent-mass term cancels, giving
\[
 \Lambda t_X(v)
 \leq O(1)\log\frac{m_p}{m_v}
 \leq O(1)\log\frac{\mu(O_v)}{\mu(I_v)}.
\]
Here the constants absorb $1/\varepsilon$, and $O_v,I_v$ are the
outer and inner balls just described.  Geometry places the source
core inside $O_v$, while $I_v$ lies in the endpoint core.  The local
entropy telescope therefore pays for every far capture, including
captures deep inside a cell.

Near captures require additional care because an inherited threshold
may greatly exceed a vertex's present personal rate.  We begin a
new run at the first capture, whenever the vertex is ineligible,
or when the threshold has increased inside its current parent.
At these \emph{seed} captures, the threshold is controlled by the
vertex's personal rate.  The first subsequent charge is at most a
constant multiple of the seed charge; later charges decrease
geometrically because eligibility contracts the scale faster than
threshold inheritance increases it.  The run's total charge is
therefore bounded by a constant times its seed charge.

A seed bound initially contains an additive constant from
$\log(2M/m_v)$.  Paying that constant once per occupied band would
destroy the desired estimate.  We prove that a seed with sufficiently
small entropy is deep inside its cell and cannot be charged for
separating an active near edge.  Its inner ball contains almost all the parent mass,
while the seed bound gives the source core a constant fraction of that
mass.  The two balls must therefore intersect.  The next distance to the
protected path is then so small that a further capture requires a
fresh threshold increase.  Such a seed cannot sponsor a later
inherited-rate charge either.  The remaining seeds have enough entropy
to absorb the additive constant.

\paragraph{Accounting for near vertices together.}
At a near seed, the rate estimate still involves the whole parent
mass.  Split its logarithm into a local inner-to-outer mass ratio and
the remaining parent-to-outer ratio.  The first part has already been
handled by nested balls.  To bound the second, consider one actual
child cell and let $N$ and $F$ be the masses of its near and far
vertices, respectively.  All near vertices lie within each other's
outer balls.  Writing $V=N+F$, their total remaining charge is bounded
using the elementary inequality
\[
 N\log(M/N)\leq V\log(M/V)+F.
\]
The first term belongs to the entropy of the parent partition and
telescopes over the recursive partition tree.  Every far vertex has
a constant lower bound on its capture charge, so the second term is
paid by the far-capture estimate.  This aggregate step is what removes
the global parent-mass term from the local analysis.

\paragraph{Boundary events and the upper endpoint of a radius.}
For analysis, a radius is drawn from a truncated exponential law on
the source's allowance interval.  The exact separation identity
consists of a capture term and a truncation remainder.  The capture
term can be restricted to vertices whose entry thresholds are near
the interval; a deep capture contributes nothing to it.
For far opportunities, the remainder is paid by source-core masses:
their normalized squares sum to at most a constant within a parent,
and the same nested-ball argument controls their carried rates across
parents.  This covers earlier potential separations even when the
actual capture occurs later.  For near edges entering in the last
exponential length of the interval, we instead charge the binary
entropy of the actual cell and its complement.  Both kinds of
remainder therefore fit within the original entropy budget.

\subsection{Exact constructions and the size of the distribution}

\paragraph{A deterministic radius from a local inequality.}
The preceding accounting also gives a deterministic weighted oracle.
At a fixed source step, let $C(r)$ bound the weighted distance cost
of edges separated by radius $r$.  The one-step exponential analysis
shows that its expectation is at most the sum of three credits:
the charge of actual captures, the binary entropy of the split, and
the far truncation remainder.  Each credit has a bound that holds
for every complete admissible execution.  It therefore suffices to
choose a radius satisfying this same inequality locally and then sum
the resulting inequalities.

Such a radius can be found exactly.  The cell, its portal, and all
edge-entry events are constant between consecutive vertex gauges.
Enumerating the rational midpoints of these intervals gives a finite
list containing a satisfactory choice.  Binary entropy is replaced
by a rational upper bound within an absolute constant factor, and
eligibility uses only the integers $K_v$.  No probability needs to be
evaluated.  A source becomes permanently protected after it is selected,
so there are at most $n-1$ source selections.  Fixed denominators for
physical distances and polynomially bounded threshold exponents give
polynomial running time for arbitrary rational input.

\paragraph{From weighted responses to an explicit distribution.}
We use the response-oracle viewpoint of
Charikar et al., R\"acke, and Andersen--Feige~\cite{CCGGP,Racke,AF}.
For either application, a tree supplies an $m$-dimensional nonnegative
payoff vector, and the weighted oracle returns a vector whose cost
against any positive price distribution is at most $\rho$.
An adaptive-step construction produces a mixture with every expected
coordinate at most $(1+\vartheta)\rho$, using
$O(m\log(2m)/\vartheta^2)$ calls.

The update size is inversely proportional to the largest coordinate
of the current response.  Thus even a response with very large
individual stretch or congestion makes only a bounded coordinate
increment.  Prices are integer powers of a rational base, with
exponents determined by rounded accumulated coordinates.  A potential
bound limits these exponents uniformly.  Normalizing the stopping
horizon by $1/\rho$ makes the call count independent of the numerical
size of $\rho$.  All probabilities and queries consequently have
polynomial encodings.  Finally, affine dependence reduces the mixture
to at most $m+1$ trees while preserving its mean.  One additional
rational linear-algebra step either removes an affine dependence or
decreases all mean coordinates together until a probability vanishes.
This gives support at most $m$ without another oracle call.

\paragraph{Why substantially smaller support can fail.}
For fundamental-cut congestion, a quadratic dependence on the number
of vertices can be necessary even when the best congestion tends to
one.  Take two sets of $k$ vertices, put a capacity-$k^2$ star inside
each (Figure~\ref{fig:heavy-stars}), and add every cross edge with capacity one.  Every spanning
tree contains a cross edge, and its fundamental cut has capacity at
least $k^2$.  A tree given probability $q_T$ therefore contributes at
least $q_Tk^2$ to some original edge's expected congestion.  Any
distribution with congestion at most $\rho$ must satisfy
\[
 q_T\leq\rho/k^2,
 \qquad |\operatorname{supp}q|\geq\lceil k^2/\rho\rceil.
\]
Nevertheless, retaining both heavy stars and choosing one cross edge
uniformly gives congestion at most $1+2(k-1)/k^2$.  Balanced sets of
cross edges attain the support lower bound for every
$\rho\geq(k^2+2k-2)/(k^2-1)$.
Thus the support requirement reflects the large congestion of an
individual sampled tree, despite the small congestion of the fixed
mixture.  It applies to the original-edge load guarantee used for
routing.  On the same family, a single fundamental-cut tree already
approximates every cut within a constant factor.  This explains why
the $O(n\log n)$ support bound for abstract metric-tree embeddings~\cite{CCGGP}
does not carry over to the stronger congestion requirement.
\section{Preliminaries and the recursive construction}
\label{sec:preliminaries}

We delete loops at the outset and allow parallel edges.
All lengths in the metric theorem are strictly positive.
Unless stated otherwise, capacities are also strictly positive;
Section~\ref{sec:capacity-extension} explains the extension to
nonnegative capacities.  For a vertex set $A$, $\delta_G(A)$ is
the set of original edges with exactly one endpoint in $A$, and
$c_G(\delta_G(A))$ is their total capacity.
All graph distances are intrinsic to the specified induced graph.
Balls are closed vertex balls.  Logarithms without a subscript are
natural.  Every tie is broken by a fixed ordering.

The weighted theorem takes a positive length $\ell_e$ and a
nonnegative demand $a_e$ on each edge.  We call $\ell$ the
\emph{physical length vector}: every edge of the output tree keeps
this length.  Put
\[
 b_e=a_e\ell_e,\qquad B=\sum_e b_e,
 \qquad C_a(T)=\sum_{e=uv}a_ed_{T,\ell}(u,v).
\]
If $B=0$, any spanning tree suffices.  For $B>0$, the proof fixes
the positive vertex measure
\[
 \mu(v)=\sum_{e\ni v}b_e+B/n.
\]
The measure is only restricted under recursion; it is never
recomputed from a child graph.  Its total mass is $3B$, and
\[
 H_0:=\sum_v\mu(v)\log\frac{3B}{\mu(v)}\leq3B\log n.
\]
The geometric lemmas do not depend on this particular measure.
The entropy argument first applies to any fixed positive measure
and is then specialized to the one above.

Each recursive instance consists of an induced connected graph $X$,
a root $x$, and a physical shortest path $P_X$ starting at $x$.
We call $P_X$ the \emph{protected path}.  The initial instance is
$G$, with an arbitrary root and the one-vertex path at that root.
Within an instance, the algorithm repeatedly removes a connected
\emph{petal} disjoint from $P_X$.  A removed petal becomes a child
instance and has one designated original \emph{portal edge}
joining it to the residual graph.  Once only $P_X$ remains, the
algorithm retains its edges and recurses on the children.
The union of these paths, recursive trees, and portal edges is the
output tree.

The next section defines petals and proves the geometric invariants
for every admissible source and radius.  The source rule and
inherited threshold are specified in
Section~\ref{sec:dyadic-entropy}; a child's threshold is twice
the threshold at the selection of its own source.
Section~\ref{sec:local-rule} gives the finite deterministic
radius rule.  Temporary lengths used to define a petal are reset
in each instance: edges on its protected path have length
$\ell_e/2$, and all other edges have length $\ell_e$.
In particular, lengths are never repeatedly halved along recursion.
\section{Protected geodesics and recursive spanning trees}
\label{sec:audited-geometry}

The decomposition preserves one physical geodesic in each recursive
instance.  Temporarily shortening this path makes it possible to attach
the remaining vertices through original graph edges while controlling
their distances to every vertex of the preserved path.  Throughout this
section, physical lengths are the original positive lengths $\ell$.
The parameters used later satisfy
\[
 0<\varepsilon\leq\frac18,\qquad
 0<\delta\leq\frac1{256},\qquad
 \chi\geq4,\qquad A_0\geq54\chi.
\]
The structural and distance bounds hold for arbitrary admissible
radii.  The scale consequences use fixed allowances proportional
to the parent distance from the protected path.

\paragraph{One recursive instance.}
Let $X$ be a connected induced graph, let $x\in X$ be its root, and let
$P$ be a physical shortest path starting at $x$.  Assign current length
$w_e=\ell_e/2$ to the edges of $P$ and $w_e=\ell_e$ to all other edges.
Every prefix of $P$ remains current-shortest: a competing route has
current length at least half its physical length.  Choose a current
shortest-path tree $S$ rooted at $x$ that contains $P$, and set
$h(v)=d_{X,w}(x,v)$.  Such a tree exists by fixing the tight predecessors
on $P$ and then choosing a tight predecessor for every remaining
vertex.  Positive lengths make this predecessor relation acyclic.

Initially the residual is $Y=X$.  We shall maintain that $Y$ is
ancestor-closed in $S$, so that every surviving root path remains and
the current root distances in $Y$ are still $h$.  To form a cell, select
$p\in Y\setminus P$.  Write $Q_p=P_S(x,p)$, let $z_p$ be its last
vertex on $P$, and set $s_p=h(p)-h(z_p)$.  Choose a radius
$0<r_p<s_p$ and put $c_p=h(p)-r_p$.  Define the entry gauge and the cell by
\begin{equation}
\label{eq:petal-entry-gauge}
 D_p(v)=h(p)-2h(v)+\min_{q\in Q_p}
       \bigl\{h(q)+2d_{Y,w}(q,v)\bigr\},
 \qquad X_p=\{v\in Y:D_p(v)<r_p\}.
\end{equation}
Let $a_p$ be the first vertex of $Q_p$ whose height exceeds $c_p$, and
let $b_p$ be its predecessor.  Remove $X_p$ from $Y$ and retain the
portal edge $a_pb_p$.  Continue until only $P$ remains.

After completing this parent decomposition, recurse on each induced
child $X[X_p]$, with root $a_p$ and protected path $P_S(a_p,p)$.  In
the child the current lengths are reset: exactly its new protected
path is halved, and every other edge receives its original physical
length.  The output consists of the child trees, all portal edges, and
the edges of $P$.  In particular, no metric edge is introduced.

\begin{lemma}[Structure of a petal]
\label{lem:petal-structure}
The gauge $D_p$ is nonnegative, satisfies $D_p(p)=0$, and is
$4$-Lipschitz in physical residual distance.  Each cell $X_p$ is
connected, contains $a_p$ and $p$, and avoids $P$; its portal endpoint
$b_p$ remains in the residual.  Removing the cell preserves ancestor
closure in $S$.  The path $P_S(a_p,p)$ is physically shortest in the
induced child.

If $R_p=\varepsilon d_{X,\ell}(p,P)$ and
$r_p\in[R_p/2,R_p]$, then the radius is admissible and the cell
contains the entire physical core $B_{Y,\ell}(p,\delta R_p)$.
\end{lemma}

\begin{proof}
For every witness $q\in Q_p$, the triangle inequality gives
\[
 h(p)+h(q)+2d_{Y,w}(q,v)-2h(v)
 \geq h(p)-h(q)\geq0.
\]
The witness $q=p$ gives $D_p(p)=0$.  Both $h$ and each residual
distance function are $1$-Lipschitz in physical residual distance.
The expression in~\eqref{eq:petal-entry-gauge}, and hence its minimum,
is therefore $4$-Lipschitz.  On the source path, ancestor closure and
the shortest-path property give
\[
 d_{Y,w}(q,v)=|h(q)-h(v)|\quad(q,v\in Q_p),
 \qquad D_p(v)=h(p)-h(v)\quad(v\in Q_p).
\]
Thus precisely the source-path vertices above the cutoff enter the
cell.  In particular, $a_p\in X_p$, $b_p\notin X_p$, and the segment
from $a_p$ to $p$ lies in $X_p$.

For a captured vertex $v$, choose a minimizing witness $q$ and define
its excess
\begin{equation}
\label{eq:petal-witness-excess}
 \eta=h(q)+d_{Y,w}(q,v)-h(v)\geq0.
 \qquad\text{Then}\qquad 2\eta<h(q)-c_p.
\end{equation}
Consequently $q$ lies on the captured segment $a_pp$.  Every vertex
of a shortest $q$--$v$ route is captured as well.  Indeed, moving from
$v$ toward $q$ decreases the distance to $q$ by exactly the traversed
length, whereas the height can decrease by at most that length.
This proves connectivity.  Along an outward edge of $S$, the height
increases by the edge's current length while any witness distance
increases by at most that length.  The gauge cannot increase.
The cell therefore contains every surviving $S$-descendant of each
of its vertices, proving ancestor closure of the new residual.

To see that $P$ is avoided, fix $v\in P$.  For a witness $q$ on $P$,
the one-dimensional distance formula gives
\[
 h(q)+2d_{Y,w}(q,v)-2h(v)\geq-h(q)\geq-h(z_p).
\]
For a witness beyond $z_p$, physical geodesicity of $P$ and
$\ell\leq2w$ give
\[
 2|h(v)-h(z_p)|
 \leq h(q)-h(z_p)+2d_{Y,w}(q,v).
\]
The same lower bound $-h(z_p)$ follows.  Hence
$D_p(v)\geq h(p)-h(z_p)=s_p>r_p$.  All witness routes inside the
cell consequently avoid the shortened edges of $P$, and their
current lengths equal their physical lengths.  The new protected
path is a subpath of a current shortest path.  Since the entire
induced child contains no shortened edge, this path is physically
shortest in the child.

Finally, the source tail from $z_p$ to $p$ has physical length $s_p$,
so $d_{X,\ell}(p,P)\leq s_p$ and $R_p<s_p$.  On the prescribed
core the Lipschitz bound gives
$D_p(v)\leq4\delta R_p<R_p/2$, proving inclusion for every allowed
radius.
\end{proof}

\begin{lemma}[Tree construction and root control]
\label{lem:root-control}
Every adaptive sequence of sources and admissible radii produces an
original-edge spanning tree $T$ containing $P$.  For each captured
vertex $v\in X_p$, there is a physical route from $v$ to $b_p$,
contained in $X_p$ except for its portal edge, of length at most
$\frac53(h(v)-h(b_p))$.  The output tree satisfies
\begin{equation}
\label{eq:petal-root-control}
 d_T(v,b_p)\leq2\bigl(h(v)-h(b_p)\bigr),
 \qquad d_T(x,v)\leq2h(v).
\end{equation}
There are at most $|X|-1$ source selections throughout the recursion
from $X$.  Moreover, each recursive tree $T_Z$ remains a connected
subtree of the final output: the final tree path between vertices of
$Z$ is exactly their path in $T_Z$.
\end{lemma}

\begin{proof}
Every child avoids the nonempty path $P$ and is therefore smaller
than its parent.  Recursion terminates by induction on the number of
vertices.  At the parent level each portal points to a later cell or
to $P$.  The child trees, portals, and $P$ consequently form a
connected graph.  If there are $k$ cells, its number of edges is
\[
 \sum_p(|X_p|-1)+k+(|V(P)|-1)=|X|-1,
\]
so it is a tree.

Fix $v\in X_p$ and its witness $q$ from
\eqref{eq:petal-witness-excess}.  Since
$d_{Y,w}(q,v)\geq h(q)-h(v)$, we have
$h(q)-h(v)\leq\eta/2$.  Combining this with the capture inequality
gives $\eta<\frac23(h(v)-c_p)$.  The route that follows the source
path from $b_p$ to $q$ and then a shortest witness route to $v$ is
physical, by Lemma~\ref{lem:petal-structure}, and has length
\[
 h(q)-h(b_p)+d_{Y,w}(q,v)
 =h(v)-h(b_p)+\eta
 \leq\frac53\bigl(h(v)-h(b_p)\bigr).
\]

For the tree bound, let $\widehat h$ be the child's current root
distance.  Its new shortening gives
\[
 \widehat h(v)
 \leq\tfrac12\bigl(h(q)-h(a_p)\bigr)+d_{Y,w}(q,v).
\]
Inductively the child tree has root distances at most $2\widehat h$.
Including the portal and using~\eqref{eq:petal-witness-excess} yields
\begin{align*}
 d_T(v,b_p)
 &\leq h(a_p)-h(b_p)+2\widehat h(v)\\
 &\leq2h(v)-h(q)-h(b_p)+2\eta\\
 &<2h(v)-c_p-h(b_p)
 \leq2\bigl(h(v)-h(b_p)\bigr).
\end{align*}
The root bound now follows backwards through the parent cells.
On $P$ it holds with equality, since the retained physical lengths
are twice the current lengths.

For the source count, a selected source becomes a vertex of its
child's protected path and can never again be selected in any
descendant.  Different branches have disjoint vertex sets, so all
selected sources are distinct initially unprotected vertices.
Finally, every recursive tree is retained as a connected subgraph
of the final tree.  Uniqueness of tree paths proves the last claim.
Several outside branches may attach to this subtree; connectivity
and acyclicity alone give the required path property.
\end{proof}

\begin{lemma}[Protection of the entire geodesic]
\label{lem:path-protection}
For every $q\in P$ and $u\in X$,
\begin{equation}
\label{eq:protected-geodesic-distance}
 d_T(q,u)\leq34d_{X,\ell}(q,u).
\end{equation}
In particular, writing $t_X(u)=d_{X,\ell}(u,P)$, every original edge
$uv$ of the instance satisfies
\begin{equation}
\label{eq:protected-edge-distance}
 d_T(u,v)\leq34\bigl(2t_X(u)+\ell_{uv}\bigr).
\end{equation}
These inequalities also hold for the corresponding paths in the
final tree of any larger instance containing this recursive call.
\end{lemma}

\begin{proof}
Follow the same-parent portal chain from $u$ until its first vertex
$b\in P$.  Put $L=h(u)-h(b)$ and $D=d_{X,\ell}(u,q)$.
The tree and physical comparison bounds of
Lemma~\ref{lem:root-control} telescope along this chain: the tree
route from $u$ to $b$ has length at most $2L$, and there is a physical
comparison route of length at most $5L/3$.  Since $h$ is
$1$-Lipschitz in physical distance and $P$ is physically geodesic,
\[
 2(L-D)\leq2|h(q)-h(b)|
 =\ell(P[q,b])\leq D+5L/3.
\]
Thus $L\leq9D$ and $\ell(P[q,b])\leq16D$.  The retained path
from $q$ to $b$ followed by the tree route to $u$ has length at most
$16D+18D=34D$.

For an edge $uv$, choose a nearest protected-path vertex $q$ to
$u$.  Then $d_{X,\ell}(q,v)\leq t_X(u)+\ell_{uv}$.
Applying~\eqref{eq:protected-geodesic-distance} to both endpoints
and using the tree triangle inequality proves
\eqref{eq:protected-edge-distance}.  The final assertion is the
connected-subtree property of Lemma~\ref{lem:root-control}.
\end{proof}

\begin{lemma}[Ambient locality and eligible contraction]
\label{lem:locality}
Suppose $p$ captures $u$.  With the parent distance
$t_X(u)=d_{X,\ell}(u,P)$ fixed before any residual removals,
\begin{align}
 h_{\rm child}(u)&\leq h(u)-c_p,
 \label{eq:child-height-cutoff}\\
 d_{X_p,\ell}(p,u)&\leq3t_X(u)-s_p+4r_p,
 \label{eq:ambient-source-locality}\\
 h_{\rm child}(u)&\leq3t_X(u)-s_p+3r_p.
 \label{eq:ambient-height-locality}
\end{align}
In particular, these bounds use the original parent distance to
$P$, not a residual distance that may increase during the carving.

Suppose further that $R_v=\varepsilon t_X(v)$ is fixed throughout
the parent, $r_p\in[R_p/2,R_p]$, and $p$ maximizes $h(v)-R_v$
over an eligible set containing $u$.  Then
\begin{equation}
\label{eq:eligible-geometric-contraction}
 h_{\rm child}(u)\leq R_u,
 \qquad d_{X_p,\ell}(p,u)\leq R_p+R_u,
 \qquad
 \frac{1-\varepsilon}{1+\varepsilon}
 \leq\frac{R_p}{R_u}\leq
 \frac{1+\varepsilon}{1-\varepsilon}.
\end{equation}
\end{lemma}

\begin{proof}
Put $A=h(u)-h(p)$ and let $q_0$ be a minimizing witness for $u$,
with excess $\eta_0$ as in~\eqref{eq:petal-witness-excess}.
The source--witness--$u$ route is entirely inside the cell and has
physical length
\[
 d_0=A+2\bigl(h(p)-h(q_0)\bigr)+\eta_0
 \leq A+2\bigl(h(p)-h(q_0)+2\eta_0\bigr)
 <A+2r_p.
\]
Choose a nearest vertex $q\in P$ to $u$ in the original parent
metric, and write $t=t_X(u)$.  Concatenate the physical source tail
$z_pp$, the preceding in-cell route, and an ambient physical
shortest $u$--$q$ path.  The latter path need not survive in $Y$;
it is used only to compare lengths in $X$.  Physical geodesicity
of $P$ and $h(q)\geq h(u)-t$ give
\[
 2(A+s_p-t)\leq2|h(q)-h(z_p)|
 \leq s_p+A+2r_p+t.
\]
Therefore $A+s_p\leq3t+2r_p$, and the bound on $d_0$ proves
\eqref{eq:ambient-source-locality}.

The same witness route, with the child's new shortening, gives
\begin{align*}
 h_{\rm child}(u)
 &\leq h(u)-\tfrac12h(q_0)-\tfrac12h(a_p)+\eta_0\\
 &<h(u)-\tfrac12c_p-\tfrac12h(a_p)
 \leq h(u)-c_p.
\end{align*}
This proves~\eqref{eq:child-height-cutoff}.  Substituting
$h(u)-c_p=A+r_p$ proves~\eqref{eq:ambient-height-locality}.

For an eligible $u$, source maximality gives
\[
 c_p\geq h(p)-R_p\geq h(u)-R_u,
 \qquad A\leq R_u-R_p.
\]
The child-height bound follows immediately, while the canonical
in-cell route has length at most
$A+2r_p\leq R_u+R_p$.  Finally, $t_X$ is $1$-Lipschitz in
the ambient physical metric, so
\[
 |R_p-R_u|\leq\varepsilon d_{X,\ell}(p,u)
 \leq\varepsilon(R_p+R_u).
\]
Rearranging yields both allowance ratios.
\end{proof}

The scale information needed by the entropy analysis follows from
these geometric bounds.  In particular, the factor-three comparison
below does not compound with recursion depth: current root heights
decrease from one instance to the next.

\begin{lemma}[Consequences for scales and boundary charges]
\label{lem:scale-corollaries}
Use fixed allowances $R_v=\varepsilon t_X(v)$, allowed radii
$r_p\in[R_p/2,R_p]$, and the eligible priority rule of
Lemma~\ref{lem:locality}.  Every capture satisfies
\begin{equation}
\label{eq:scale-geometric-inputs}
 d_{Y,\ell}(p,u)\leq3t_X(u),\qquad
 t_X(p)<3t_X(u),\qquad
 h_{\rm child}(u)\leq\min\{h(u),3t_X(u)\},\qquad
 t_X(u)\leq h(u).
\end{equation}
Along a vertex's recursive history, let $t_i$ be its parent distance
to the protected path and $\sigma_i=\min_{j\leq i}t_j$.  Then
$\sigma_i\leq t_i\leq3\sigma_i$.

A capture with $R_p\leq R_u/\chi$ must have an ineligible endpoint
$u$.  For the near set $N_p=\{u\in X_p:R_u<\chi R_p\}$,
every two vertices are at physical distance less than
$6\chi t_X(p)$ in the same residual $Y$.  If $\sigma_X(u)$ is
the running minimum at this parent, then
\begin{equation}
\label{eq:near-set-outer-ball}
 N_p\subseteq B_{Y,\ell}(u,A_0\sigma_X(u))
 \qquad(u\in N_p).
\end{equation}
Moreover, any physical shortest $p$--$u$ path in $Y$ of length
less than $R_p/8$ lies in every allowed cell, and bounds the child's
distance from $u$ to its new protected path.

For an edge of length $\ell$, stop tracking it in an instance as
soon as an endpoint has $t_X\leq16\chi\ell/\varepsilon$.
Its final tree distance is then at most
$34(32\chi/\varepsilon+1)\ell$.
While both endpoints remain above this threshold, either endpoint
$u$ gives the bound $d_T(u,v)\leq Wt_X(u)$ with $W=3\cdot34=102$.
If $R_u<\chi R_p$ for one endpoint and the source can separate
such an edge by an allowed radius, its lower entry threshold is
greater than $R_p/4$.
\end{lemma}

\begin{proof}
The allowance satisfies $r_p\leq\varepsilon t_X(p)\leq
\varepsilon s_p$.  Since $\varepsilon\leq1/8$, the two locality
bounds in Lemma~\ref{lem:locality} give
$d_{Y,\ell}(p,u)\leq3t_X(u)$ and
$h_{\rm child}(u)\leq3t_X(u)$.  The cutoff is positive, so
\eqref{eq:child-height-cutoff} also gives
$h_{\rm child}(u)\leq h(u)$.  The suffix of the $S$-root path to
$u$ after its last vertex on $P$ is physical and has length at most
$h(u)$, proving $t_X(u)\leq h(u)$.  Physical Lipschitzness and the
more precise source-locality bound yield
\[
 t_X(p)\leq t_X(u)+d_{X,\ell}(p,u)
 \leq4t_X(u)-t_X(p)+4\varepsilon t_X(p).
\]
Consequently
$t_X(p)\leq2t_X(u)/(1-2\varepsilon)\leq8t_X(u)/3<3t_X(u)$.
For an earlier index $j<i$, decreasing root heights give
$t_i\leq h_i\leq h_{j+1}\leq3t_j$.  Taking the running minimum
proves the claimed comparison with $\sigma_i$.

An eligible captured endpoint would satisfy
$R_p/R_u\geq(1-\varepsilon)/(1+\varepsilon)\geq7/9>1/\chi$,
excluding a far capture.  For $u\in N_p$ we have
$t_X(p)/3<t_X(u)<\chi t_X(p)$, and hence
$\sigma_X(u)>t_X(p)/9$.  Joining two near vertices through the
source gives residual distance less than $6\chi t_X(p)$.
Since $A_0\geq54\chi$, their outer balls contain the full near
set, proving~\eqref{eq:near-set-outer-ball}.
For a shortest residual path of length less than $R_p/8$, every
vertex on the path has gauge below $4(R_p/8)=R_p/2$ and is
therefore captured at every allowed radius.  The source belongs to
the new protected path, so this retained physical path bounds the
child's distance to that path.

The stopping and active bounds follow from
\eqref{eq:protected-edge-distance}.  At a stop use the endpoint
with $t_X\leq16\chi\ell/\varepsilon$; during activity,
$\ell<t_X(u)$ and $34(2t_X(u)+\ell)\leq102t_X(u)$.
If an active endpoint is near, then
\[
 R_p>R_u/\chi=\varepsilon t_X(u)/\chi>16\ell.
\]
The two entry gauges differ by at most $4\ell$.  If an allowed
radius can separate the edge, the larger gauge is at least
$R_p/2$, so the smaller is at least
$R_p/2-4\ell>R_p/4$, as claimed.
\end{proof}

\paragraph{Discrete choices and hypothetical captures.}
All conclusions hold without uniqueness of shortest paths, edge
subdivision, or perturbation of lengths.  A cutoff equal to a
path-vertex height excludes that vertex and uses the first vertex
of greater height as $a_p$; any minimizing gauge witness gives
the same inequalities.  All child graphs are induced and retain
every original edge whose endpoints survive.  The source, the
eligible set, the residual, and the parent data are fixed before
the radius is chosen.  The lemmas therefore also apply to any
hypothetical allowed-radius capture at that state, independently
of the numerical rule that defines eligibility.
\section{Entropy bounds for the recursive construction}
\label{sec:dyadic-entropy}

Fix positive rational constants $\varepsilon,\delta,\chi,A_0,\zeta$
satisfying
\begin{equation}
\label{eq:admissible-parameters}
 0<\varepsilon\leq\frac18,\qquad
 0<\delta\leq\frac1{256},\qquad
 \chi\geq4,\qquad A_0\geq54\chi,\qquad
 0<\zeta\leq2^{-10}.
\end{equation}
The parameters control the local allowance, core radius, far-source
separation, outer-ball radius, and low-entropy cutoff, respectively.
They are absolute constants, independent of the input.  We give one
concrete choice in Section~\ref{sec:choice-of-parameters}.
Throughout this section the physical lengths and the positive vertex
measure $\mu$ remain fixed.  Write
\[
 M_0=\mu(V),\qquad H_0=\sum_{v\in V}\mu(v)\log\frac{M_0}{\mu(v)}.
\]
All unmarked logarithms are natural.

\subsection{Rates and charges}

\paragraph{The source rule.}
In a parent $X$ with protected path $P_X$, keep
\[
 t_X(v)=d_{X,\ell}(v,P_X),\qquad R_v=\varepsilon t_X(v),\qquad
 L_v=h(v)-R_v,\qquad M=\mu(X)
\]
fixed throughout its carving.  The candidates are the surviving vertices
outside $P_X$.  In residual $Y$, define
\[
 C_v=B_{Y,\ell}(v,\delta R_v),\qquad m_v=\mu(C_v),\qquad
 \kappa_v=\log\frac{2M}{m_v},\qquad
 K_v=\left\lceil\log_2\frac{2M}{m_v}\right\rceil.
\]
The integer $K_v$ is computed by comparisons with powers of two.
A candidate is eligible when $4K_v/R_v\leq\Lambda$; among eligible
candidates choose a source maximizing $L_v$, with a fixed tie rule.
The root threshold is the minimum initial candidate rate.  A threshold
doubles only when no candidate is eligible, and a child inherits twice
the threshold at its own source's selection.  The radius may be any
$r_p\in(R_p/2,R_p)$.
Since $\kappa_v\geq\log2$,
\begin{equation}
\label{eq:dyad-log-comparison}
 \kappa_v\leq K_v\leq\frac{\kappa_v}{\log2}+1
       \leq\frac{2\kappa_v}{\log2}<3\kappa_v.
\end{equation}
Thus source eligibility implies $\Lambda R_p\geq4\log(2M/m_p)$.
Within one parent, residual restriction can only decrease $m_v$, so
$K_v$ is nondecreasing.

\paragraph{Geometric consequences.}
Lemmas~\ref{lem:locality} and~\ref{lem:scale-corollaries} imply, for every capture of $v$ by $p$,
\begin{equation}
\label{eq:dyad-arbitrary-geometry}
 d_{Y,\ell}(p,v)\leq3t_X(v),\quad t_X(p)<3t_X(v),\quad
 h_{\rm child}(v)\leq\min\{h(v),3t_X(v)\},\quad t_X(v)\leq h(v).
\end{equation}
Indeed $\varepsilon\leq1/8$ gives
$t_X(p)\leq2t_X(v)/(1-2\varepsilon)\leq8t_X(v)/3<3t_X(v)$.
If the captured vertex is eligible, then
\begin{equation}
\label{eq:dyad-eligible-geometry}
 h_{\rm child}(v)\leq\varepsilon t_X(v),\qquad
 \frac{1-\varepsilon}{1+\varepsilon}
 \leq\frac{R_p}{R_v}\leq
 \frac{1+\varepsilon}{1-\varepsilon}.
\end{equation}
These conclusions also hold for a hypothetical allowed-radius capture
at the unchanged pre-draw state.  The entry gauge $D_p$ is nonnegative,
$4$-Lipschitz in physical residual distance, and zero at $p$.
Consequently $C_p$ lies in every allowed cell, since
$4\delta R_p\leq R_p/64<R_p/2$.
If $d_{Y,\ell}(p,v)<R_p/8$, the entire physical shortest $p$--$v$
path lies in every allowed cell.  Since $p$ is on the child's protected
path, this also gives $t_{\rm child}(v)\leq d_{Y,\ell}(p,v)$.

Follow a vertex's actual capturing cells until it reaches a protected
path.  At its $i$th capture, write $t_i,\Lambda_i,Y_i,M_i$ for the
parent scale, capture threshold, pre-capture residual, and parent mass.
Put
\[
 \sigma_i=\min_{j\leq i}t_j,\qquad x_i=\Lambda_it_i,\qquad
 O_i=B_{Y_i,\ell}(v,A_0\sigma_i),\qquad
 I_i=B_{Y_i,\ell}(v,\delta\varepsilon\sigma_i),\qquad
 B_i=\log\frac{M_i}{\mu(I_i)}.
\]
A capture is \emph{far} if $R_p\leq R_v/\chi$ and \emph{flagged}
if $D_p(v)>R_p/4$.  Define the total charges
\[
 Q_F=\sum_v\mu(v)\sum_{i:\,\mathrm{far}}x_i,
 \qquad Q_M=\sum_v\mu(v)\sum_{i:\,\mathrm{flagged}}x_i.
\]

\begin{theorem}[Main-charge bound]
\label{thm:dyad-main}
For every admissible sequence of radii,
\[
 Q_F\leq C_FH_0,\qquad Q_M\leq C_MH_0,
\]
where
\begin{align*}
 L&=1+\left\lceil\log_2\frac{A_0}{\delta\varepsilon}\right\rceil,
 &C_F&=\frac{144\chi L}{(\chi-3)\varepsilon},\\
 C_R&=1+\frac6{1-2\varepsilon},
 &C_S&=\frac{24}{\varepsilon}
                       \left(1+\frac{\log2}{\zeta}\right),\\
 C_M&=C_R\left[C_F+12C_S
              \left(L+1+\frac{\varepsilon C_F}{4\chi}\right)\right].
\end{align*}
These constants depend only on the parameters in
\eqref{eq:admissible-parameters}.
\end{theorem}

\subsection{Far captures and scale telescoping}

\begin{lemma}[Running minima and occupied bands]
\label{lem:dyad-bands}
Along every vertex history,
$\sigma_i\leq t_i\leq3\sigma_i$, the residuals $Y_i$ are nested,
and $\Lambda_{i+1}\geq2\Lambda_i$.
For any subset $J$ of capture indices, retain its last index $j_r$ in
each occupied dyadic band of $\sigma_i$, ordered chronologically.  Then
\[
 \sum_{i\in J}x_i\leq12\sum_r x_{j_r},\qquad
 \sum_r\log\frac{\mu(O_{j_r})}{\mu(I_{j_r})}
       \leq L\log\frac{M_0}{\mu(v)}.
\]
\end{lemma}
\begin{proof}
If $j<i$ realizes the running minimum, monotonicity of the root heights
and \eqref{eq:dyad-arbitrary-geometry} give
$t_i\leq h_i\leq h_{j+1}\leq3t_j$; the case $j=i$ is immediate.
The next parent is the actual cell capturing the vertex, so its next
pre-capture residual is contained in $Y_i$.  Its threshold starts at
$2\Lambda_i$ and can only increase.

Within one dyadic band, the scales $t_i$ differ by less than a factor
six.  Backwards from the last selected index, the rates decrease by
factors at least two.  Summing this geometric series proves the first
inequality.  Successive occupied bands have strictly decreasing integer
indices, so $\sigma_{j_{r+L}}<2^{1-L}\sigma_{j_r}$.
The definition of $L$ gives
$A_0\sigma_{j_{r+L}}\leq\delta\varepsilon\sigma_{j_r}$.
Physical distances cannot decrease under induced restriction; hence
$O_{j_{r+L}}\subseteq I_{j_r}$.
For each residue class modulo $L$, the logarithmic mass ratios telescope
to at most $\log(M_0/\mu(v))$, since every inner ball contains $v$.
Summing these bounds proves the second inequality.
\end{proof}

\begin{lemma}[Far captures]
\label{lem:dyad-far}
For every vertex,
\[
 \sum_{i:\,\mathrm{far}}x_i
       \leq C_F\log\frac{M_0}{\mu(v)}.
\]
This estimate includes far captures that are not flagged.
\end{lemma}
\begin{proof}
An eligible captured vertex satisfies
$R_p/R_v\geq(1-\varepsilon)/(1+\varepsilon)\geq7/9>1/\chi$.
Thus every far captured vertex is ineligible.  By
\eqref{eq:dyad-log-comparison}, endpoint ineligibility and source
eligibility imply
\[
 \kappa_v>\frac{\Lambda R_v}{12},\qquad
 \kappa_p\leq K_p\leq\frac{\Lambda R_p}{4}
                         \leq\frac{\Lambda R_v}{4\chi}.
\]
The fixed parent numerator cancels.  Writing
$g_\chi=(\chi-3)/(12\chi)>0$, we obtain
\begin{equation}
\label{eq:dyad-local-gap}
 \log\frac{m_p}{m_v}>g_\chi\Lambda R_v
                              =g_\chi\varepsilon x_i.
\end{equation}
The source core lies in $O_i$: its center is within $3t_i$ of $v$,
its radius is at most $\delta\varepsilon t_i/\chi$, and
$t_i\leq3\sigma_i$.  The parameter bounds give
$3(3+\delta\varepsilon/\chi)<10<A_0$.
The endpoint core contains $I_i$.  Therefore
\begin{equation}
\label{eq:dyad-local-certificate}
 x_i<\frac1{\varepsilon g_\chi}
                 \log\frac{\mu(O_i)}{\mu(I_i)}.
\end{equation}
Apply Lemma~\ref{lem:dyad-bands} first to all far indices and then to
their representatives.  The resulting constant is
$12L/(\varepsilon g_\chi)=C_F$.
\end{proof}

\subsection{Near captures and the seed argument}

\begin{lemma}[Seed reduction]
\label{lem:dyad-seeds}
Call a capture a \emph{seed} if it is the vertex's first capture, if the
vertex is ineligible, or if the threshold increased inside its parent
before the capture.  Every seed satisfies
\[
 x_i\leq\frac8\varepsilon K_v
          \leq\frac{24}\varepsilon(\log2+B_i).
\]
If $B_i\leq\zeta$, the seed is unflagged and its next capture, if
any, follows an internal threshold increase.  Consequently
\[
 \sum_{i:\,\mathrm{flagged}}x_i
 \leq C_R\sum_{\substack{i:\,\mathrm{seed}\\B_i>\zeta}}x_i,
 \qquad x_i\leq C_SB_i
       \quad\text{at each seed in the latter sum}.
\]
\end{lemma}
\begin{proof}
At an ineligible seed, $\Lambda R_v<4K_v$.  Otherwise consider the
last local doubling before the capture.  Immediately before it, the
vertex was still present and ineligible.  Monotonicity of $K_v$ gives
$\Lambda R_v<8K_v$ at capture.  At the initial root, if there was no
increase, the initial minimum rate is at most the vertex's current rate.
This proves the first bound.  The endpoint core contains $I_i$, so
$\kappa_v\leq\log2+B_i$; now apply \eqref{eq:dyad-log-comparison}.

Suppose $B_i\leq\zeta$.  Since
$2M_i/m_v\leq2e^{B_i}\leq2e^\zeta<4$, one has $K_v\leq2$ and
$\Lambda R_v\leq16$.  The strict source-scale bound $R_p<3R_v$
and source eligibility yield $K_p\leq\Lambda R_p/4<12$.
Integrality gives $K_p\leq11$, and consequently
\[
 m_p\geq\frac{2M_i}{2^{11}}=2^{-10}M_i.
\]
The complement of $I_i$ in its parent has mass
$M_i(1-e^{-B_i})<2^{-10}M_i$: for $B_i>0$ use
$1-e^{-B_i}<B_i\leq\zeta\leq2^{-10}$, and for $B_i=0$ the
complement has mass zero.  Thus the source core intersects $I_i$.
Also $\Lambda R_p\geq4K_p\geq4$ gives $R_p\geq R_v/4$.
The intersection supplies
\[
 d_{Y_i,\ell}(p,v)\leq\delta(R_p+R_v)<4\delta R_v
                                      \leq16\delta R_p.
\]
Every vertex on a physical shortest $p$--$v$ path has entry gauge less
than $64\delta R_p\leq R_p/4$.  The seed is therefore unflagged.
The entire path is inside every allowed cell, and the source is on the
new protected path, so $t_{i+1}\leq4\delta\varepsilon t_i$.

Every actual capture satisfies the floor
$\Lambda t_X(v)>4/(3\varepsilon)$, because
$R_p<3\varepsilon t_X(v)$ and $\Lambda R_p\geq4$.
If the next capture used only the inherited threshold, then
\[
 x_{i+1}\leq8\delta\varepsilon x_i
       =8\delta\Lambda R_v\leq128\delta
       \leq\frac12<\frac4{3\varepsilon},
\]
a contradiction.  Thus the next capture, if present, follows an internal
threshold increase and is another seed.

Every maximal nonseed run follows a seed.  Its first charge is at most
six times that seed charge, by threshold inheritance and the arbitrary
scale reset.  Every subsequent predecessor is eligible, so later
charges contract by $2\varepsilon<1$.
The entire run costs at most $6/(1-2\varepsilon)$ times its preceding
seed.  A low-entropy seed is unflagged and cannot precede a nonseed run.
Assigning the remaining flagged charges to high-entropy seeds proves
the factor $C_R$.  At these seeds,
$\log2+B_i\leq(1+\log2/\zeta)B_i$, proving the bound with $C_S$.
\end{proof}

\begin{proof}[Proof of Theorem~\ref{thm:dyad-main}]
Multiply Lemma~\ref{lem:dyad-far} by $\mu(v)$ and sum to obtain the
far bound.  For flagged charges apply Lemma~\ref{lem:dyad-seeds}.
Far high-entropy seeds are already paid by $Q_F$.
For each vertex, group its remaining near high-entropy seeds into
occupied dyadic bands and retain the last such seed in each band.
Lemma~\ref{lem:dyad-bands} reduces their total charge to twelve times
the sum over representatives.  At each representative write
\[
 B_i=\log\frac{\mu(O_i)}{\mu(I_i)}
                   +\log\frac{M_i}{\mu(O_i)}.
\]
The first terms, summed with vertex masses, contribute at most $LH_0$.

For the second terms fix an actual source cell $X_p$ in a parent of
mass $M$ and partition it as
\[
 N_p=\{v\in X_p:R_v<\chi R_p\},\qquad F_p=X_p\setminus N_p.
\]
Write $N=\mu(N_p)$, $F=\mu(F_p)$, and $V=N+F$.
Every near representative in this cell belongs to $N_p$.
For $u,v\in N_p$, \eqref{eq:dyad-arbitrary-geometry} gives
\[
 t_X(p)/3<t_X(u),t_X(v)<\chi t_X(p),\qquad
 d_{Y,\ell}(u,v)<6\chi t_X(p).
\]
Since $\sigma(u)\geq t_X(u)/3>t_X(p)/9$ and $A_0\geq54\chi$,
its outer ball contains all of $N_p$.
Thus, for any representative subset of $N_p$,
\begin{align*}
 \sum_{u\text{ represented in }N_p}\mu(u)
               \log\frac M{\mu(O(u))}
 &\leq N\log(M/N)\\
 &\leq V\log(M/V)+F.
\end{align*}
The last inequality follows because the derivative of $z\log(M/z)$
is at least $-1$ on $(0,M]$.
The terms $V\log(M/V)$ sum to at most $H_0$ over the recursive
partition: include the retained protected path as a terminal part of
each parent partition and telescope the complete partition entropies.
For $v\in F_p$,
\[
 \Lambda t_X(v)\geq\frac{\chi\Lambda R_p}{\varepsilon}
                    \geq\frac{4\chi}{\varepsilon}.
\]
Therefore $\sum_{\mathrm{cells}}F\leq\varepsilon Q_F/(4\chi)$.
Combining the local mass ratios, the global mass cancellation, the band
factor twelve, and the seed reduction gives the stated $C_M$.
\end{proof}

\subsection{Credits for missed far opportunities}

\begin{theorem}[Far-opportunity credits]
\label{thm:dyad-tail}
Fix an original edge $e=uv$.  Follow its residual carving until first
capture in each parent.  Continue to a child only if both endpoints
enter it, and stop at separation, a protected-path endpoint, or an
earlier analytical stopping rule.  At a reached source step set
\[
 \tau=\min\{D_p(u),D_p(v)\},\qquad
 \omega=\max\{D_p(u),D_p(v)\},\qquad
 I=\mathbf1_{\{\max(\tau,R_p/2)<\min(\omega,R_p)\}}.
\]
Label the step by an endpoint $z$ attaining $\tau$, with a fixed tie
rule.  It is far-labelled $z$ if $I=1$ and $R_p\leq R_z/\chi$.
Let $\theta_j$ be any nonnegative pre-choice weights satisfying
\[
 \sum_{j\text{ reached in one parent}}\theta_j\leq\frac13
\]
for every realized history.  Then, separately for each fixed endpoint
label,
\[
 \sum_X t_X(z)\sum_{j\text{ far-labelled }z}\Lambda_j\theta_j
 \leq C_T\log\frac{M_0}{\mu(z)},\qquad
 C_T=\frac{48\chi L}{(\chi-3)\varepsilon}.
\]
\end{theorem}
\begin{proof}
Boundary overlap implies that some allowed radius captures $z$.
If $z$ were eligible, \eqref{eq:dyad-eligible-geometry} for that
hypothetical capture would contradict $R_p\leq R_z/\chi$.
Thus $z$ is ineligible, and the local gap
\eqref{eq:dyad-local-gap} applies even if the actual radius misses both
endpoints.  Hypothetical capture also gives
$d_{Y,\ell}(p,z)\leq3t_X(z)$, so the source core lies in the local
outer ball and the inner ball lies in the endpoint core.
Use the running minimum over the entire surviving vertex history,
including parents omitted below.  The local certificate
\eqref{eq:dyad-local-certificate} therefore holds at every such step.

In each parent with far-labelled steps, retain the last reached one,
with threshold $\overline\Lambda_X^z$ and its pre-choice residual.
Monotonicity of the threshold and the packing hypothesis imply
\[
 t_X(z)\sum_{j\text{ far-labelled }z}\Lambda_j\theta_j
       \leq\frac13\overline\Lambda_X^z t_X(z).
\]
These retained residuals are nested: the actual first-capturing cell
lies in the residual at the retained step, and each subsequent residual
lies inside that child.  Their rates grow by at least two, because a
child starts at twice the actual first-capture threshold, which is at
least the preceding retained threshold.  Omitting intervening parents
preserves both properties.

Apply the occupied-band argument to these retained parents.  Within
one band the scales differ by less than six, and the rates form a
backwards geometric series, giving factor twelve.  At the last
representative of each band the local mass certificate applies.
After $L$ occupied bands the later outer ball lies inside the earlier
inner ball.  The logarithms thus telescope to at most
$L\log(M_0/\mu(z))$.  The product of constants is
$(1/3)\cdot12L/(\varepsilon g_\chi)=C_T$.
\end{proof}

\begin{corollary}[Rational packed credits]
\label{cor:dyad-rational-tail}
Set $\theta_p=m_p^2/(3M^2)$ at a source step.
These weights satisfy Theorem~\ref{thm:dyad-tail} for every admissible
radius sequence.  If $b_e\geq0$ and
$\mu(v)\geq\sum_{e\ni v}b_e$, then
\[
 \sum_e b_e\sum_{\text{active far steps of }e}
                  t_X(z_e)\Lambda_p\theta_p\leq C_TH_0.
\]
\end{corollary}
\begin{proof}
Guaranteed cores lie in their actual removed cells, which are disjoint
in a parent.  Hence
\[
 \sum_{p\text{ reached}}\theta_p
 =\frac13\sum_p(m_p/M)^2
 \leq\frac13\left(\sum_pm_p/M\right)^2\leq\frac13.
\]
Apply Theorem~\ref{thm:dyad-tail} separately to each fixed endpoint
label of every edge.  After multiplying by $b_e$ and summing,
\[
 \sum_{e=uv}b_e
 \left(\log\frac{M_0}{\mu(u)}+\log\frac{M_0}{\mu(v)}\right)
 \leq\sum_v\mu(v)\log\frac{M_0}{\mu(v)}=H_0.
\]
Omitting stopped or separated edges only removes nonnegative terms.
\end{proof}
\section{A deterministic radius rule}
\label{sec:local-rule}

We now construct a weighted low-stretch tree.  Fix the parameters from
\eqref{eq:admissible-parameters}, and let
\[
 C_{\rm geo}=34,\qquad W=3C_{\rm geo}.
\]
The constant $C_{\rm geo}$ is the protected-path guarantee in
Lemma~\ref{lem:path-protection}; $W$ bounds the cost of separating
an edge whose endpoint scales exceed its length.
An auxiliary exponential density is used only to establish a local
averaging inequality.  The algorithm chooses a rational radius
satisfying that inequality.

\subsection{Fixed measure and active edges}

For nonnegative demands $a_e$ on the original edges, put
$b_e=a_e\ell_e$ and $B=\sum_e b_e$.
If $B=0$, return any spanning tree.  Otherwise use the fixed measure
\begin{equation}
\label{eq:fixed-measure}
 \mu(v)=\sum_{e\ni v}b_e+\frac Bn,\qquad
 M_0=3B,\qquad
 H_0=\sum_v\mu(v)\log\frac{M_0}{\mu(v)}\leq3B\log n.
\end{equation}
Start with a singleton protected path at an arbitrary root, and apply
the source rule of Section~\ref{sec:dyadic-entropy}.
Complete each parent carving before processing its children in source
order.  Each child stores twice its own source-time threshold, regardless
of subsequent increases in its parent.

Maintain a list of active original demand edges.  At entry to a parent
$X$, stop the analysis of an edge $e=uv$ when
\begin{equation}
\label{eq:dyad-stop}
 \min\{t_X(u),t_X(v)\}\leq
                         \frac{16\chi}{\varepsilon}\ell_e.
\end{equation}
Protected-path geometry and the connected-subtree invariant bound its
final tree distance by $K_0\ell_e$, where
\[
 K_0=C_{\rm geo}\left(\frac{32\chi}{\varepsilon}+1\right).
\]
Pay this cost once.  An active edge separated by a selected cell is
also closed.  An edge whose endpoints both enter a cell passes to the
child; an edge whose endpoints both remain stays in the residual list.
Stopping or closing an edge changes only this list, not the graph,
measure, or source rule.  Complete the structural construction even
when the active list is empty.

For an active edge at a source step, either endpoint $z$ gives the
pre-choice bound
\begin{equation}
\label{eq:dyad-predraw-cost}
 d_T(u,v)\leq C_{\rm geo}(2t_X(z)+\ell_e)\leq Wt_X(z).
\end{equation}
Here $t_X(z)>\ell_e$ follows from the stopping rule and
\eqref{eq:admissible-parameters}.  These estimates hold in the final
global tree because each recursively constructed tree is a connected
subtree of it.

\paragraph{Implementation of the weighted construction.}
Algorithm~\ref{alg:weighted-tree} specifies the order of all graph
operations.  The original graph, physical lengths, demands, baseline
weights, and vertex measure are shared by its recursive calls.  Only
the temporary lengths, distance tables, residual vertex sets, and
active-edge lists belong to an individual instance.  An active-edge
list is bookkeeping for the radius objective: deleting an edge from
this list never deletes it from an induced graph.

\begin{paperalgorithm}{\textsc{WeightedTree} and its recursive procedure}
 {alg:weighted-tree}
\textbf{Input:} A connected loopless graph $G=(V,E)$, rational lengths
$\ell_e>0$, and rational demands $a_e\geq0$.  Use the parameters
of~\eqref{eq:admissible-parameters}, with the concrete choice in
Section~\ref{sec:choice-of-parameters}, and $W=102$.

\textbf{Output:} An original-edge spanning tree satisfying
Theorem~\ref{thm:dyad-weighted}.

\textbf{Procedure} $\textsc{WeightedTree}(G,\ell,a)$.
\begin{enumerate}[label=\arabic*.,leftmargin=2em,itemsep=2pt,topsep=2pt]
\item Compute $b_e=a_e\ell_e$ and $B=\sum_e b_e$.
If $B=0$, return any spanning tree, including the empty tree when
$|V|=1$.
\item Fix $\mu(v)=\sum_{e\ni v}b_e+B/|V|$ for the entire call.
Choose the first vertex $x$ in the fixed order and return
$\textsc{Build}(G,x,(x),\bot,E)$.
The symbol $\bot$ requests initialization of the root threshold.
\end{enumerate}

\textbf{Procedure} $\textsc{Build}(X,x,P,\Lambda_{\rm in},\mathcal A)$.
At entry, $X=G[V(X)]$ is connected and $P$ is a physical shortest
path in $X$ starting at $x$.  Both endpoints of every edge in
$\mathcal A$ belong to $X$.
\begin{enumerate}[label=\arabic*.,leftmargin=2em,itemsep=2pt,topsep=2pt]
\item If $V(X)=V(P)$, stop all edges in $\mathcal A$ and return
$E(P)$; their endpoint scales are zero.
\item Set $w_e=\ell_e/2$ on $E(P)$ and $w_e=\ell_e$ otherwise.
Compute current root distances $h=d_{X,w}(x,\cdot)$ and a current
shortest-path tree $S$ containing $P$: fix the predecessors on $P$
and choose a tight predecessor for each other vertex, with fixed ties.
Compute
$t(v)=d_{X,\ell}(v,P)$ by a physical multi-source shortest-path
computation, and set $R_v=\varepsilon t(v)$ and $M=\mu(V(X))$.
Keep $h,S,t,R,M$ fixed throughout this parent.
\item Remove from $\mathcal A$ each $e=uv$ with
$\min\{t(u),t(v)\}\leq16\chi\ell_e/\varepsilon$.
Initialize $Y=X$, $\Lambda=\Lambda_{\rm in}$, an empty ordered
child list, and $T=E(P)$.
\item \textbf{While $V(Y)\setminus V(P)\neq\varnothing$:}
\begin{enumerate}[label=(\alph*),leftmargin=2em,itemsep=2pt,topsep=2pt]
\item Compute all-pairs distances in $(Y,\ell)$ and $(Y,w)$.
For each candidate $v\in V(Y)\setminus V(P)$ compute
\[
 m_v=\sum_{u:\,d_{Y,\ell}(v,u)\leq\delta R_v}\mu(u),
 \quad K_v=\min\{k\in\mathbb Z_{\geq0}:2^k m_v\geq2M\},
 \quad \lambda_v=4K_v/R_v.
\]
\item If $\Lambda=\bot$, set $\Lambda=\min_v\lambda_v$.
If no $v$ satisfies $\lambda_v\leq\Lambda$, multiply $\Lambda$
by the smallest power of two that makes one candidate eligible.
Choose an eligible $p$ maximizing $h(p)-R_p$, using the fixed tie rule.
\item Call $\textsc{ChooseRadius}$ at this state to obtain
$(r,Z)$, where $Z=\{v\in Y:D_p(v)<r\}$.
Put $c=h(p)-r$.  On $P_S(x,p)$ let $a_p$ be the first vertex
with $h(a_p)>c$, and let $b_p$ be its predecessor.
\item Append the child record
\[
 \bigl(G[Z],a_p,P_S(a_p,p),2\Lambda,
       \{e=uv\in\mathcal A:u,v\in Z\}\bigr).
\]
Add the original edge $a_pb_p$ to $T$.
Close every active edge with exactly one endpoint in $Z$; remove
the child's active edges from the parent list.  Equivalently, replace
$\mathcal A$ by $\{e=uv\in\mathcal A:u,v\in V(Y)\setminus Z\}$.
Replace $Y$ by the induced graph $Y[V(Y)\setminus Z]$.
\end{enumerate}
\item Process the stored child records in source order, calling
$\textsc{Build}$ on each record and adding its returned edges to $T$.
Return $T$.
\end{enumerate}
\end{paperalgorithm}

The threshold stored for a child is twice the value at its own source
selection; later increases in the parent do not change that record.
The loop continues even when $\mathcal A$ is empty.  Thus every
vertex is included and every structural edge required by
Lemma~\ref{lem:root-control} is retained.  Repeated threshold doublings
can be batched because the residual and personal rates do not change
during those doublings.  All shortest-path computations use exact
rational arithmetic; the all-pairs implementation is the conservative
one analyzed in Theorem~\ref{thm:weighted-running-time}.

\subsection{The one-state boundary identity}

Fix the state before choosing a radius: the residual $Y$, source $p$,
allowance $R=R_p$, parent mass $M$, core mass $m=m_p$, and threshold
$\Lambda$.  Source eligibility gives
$\Lambda R\geq4\log(2M/m)$.  For analysis only, consider
\[
 f(r)=\frac{\Lambda e^{-\Lambda(r-R/2)}}{1-e^{-\Lambda R/2}},
 \qquad R/2\leq r\leq R,
 \qquad \beta=(e^{\Lambda R/2}-1)^{-1}.
\]
For an active edge $e=uv$, put
\[
 \tau_e=\min\{D_p(u),D_p(v)\},\qquad
 \omega_e=\max\{D_p(u),D_p(v)\},\qquad
 a_e^*=\max\{\tau_e,R/2\},
\]
\[
 w_e=(\min\{\omega_e,R\}-a_e^*)_+,\qquad
 I_e=\mathbf1_{\{w_e>0\}},\qquad
 \kappa_e=1-e^{-\Lambda w_e}.
\]
Choose an endpoint $z_e$ attaining $\tau_e$ by a fixed tie rule.
This label is determined before the radius and is captured whenever
the step captures either endpoint.

\begin{lemma}[Boundary identity]
\label{lem:local-boundary}
Conditional on the fixed state,
\[
 P_{{\rm cut},e}=\kappa_e(P_{{\rm cap},e}+\beta),\qquad
 \kappa_e\leq4\ell_e\Lambda I_e,
\]
where $P_{{\rm cap},e}$ is the probability that at least one endpoint
enters the selected cell.  If $I_e=1$ and
$R-a_e^*\geq1/\Lambda$, then
\[
 P_{{\rm cut},e}\leq(1-e^{-1})^{-1}\kappa_e P_{{\rm cap},e}
                  <2\kappa_e P_{{\rm cap},e}.
\]
\end{lemma}
\begin{proof}
The gauge is $4$-Lipschitz, so $w_e\leq4\ell_e$.
For $I_e=1$, direct integration gives
\[
 P_{{\rm cap},e}=\beta(e^{\Lambda(R-a_e^*)}-1),\qquad
 P_{{\rm cut},e}=\kappa_e\beta e^{\Lambda(R-a_e^*)}.
\]
The formulas include $\tau_e\leq R/2$, when the capture probability
is one.  If $I_e=0$, the cut probability and $\kappa_e$ are both zero.
The first assertion now follows from $1-e^{-x}\leq x$.
In the stated nonterminal case,
$\beta\leq P_{{\rm cap},e}/(e-1)$, proving the last inequality.
\end{proof}

A boundary-capable edge is \emph{far} at this step if
$R\leq R_{z_e}/\chi$, and \emph{near} otherwise.
A near edge is \emph{terminal} if
$R-1/\Lambda<\tau_e<R$; the other near boundary-capable edges are
nonterminal.  Since $\Lambda R\geq4$, a near nonterminal edge
satisfies $R-a_e^*\geq1/\Lambda$.
Let $\mathcal F$ be the set of far edges and $\mathcal M$ the union
of far and near nonterminal edges.  Both sets are fixed before choosing
the radius.

\subsection{Terminal windows and split entropy}

For $0\leq V\leq M$, define
\[
 H_M(V)=V\log(M/V)+(M-V)\log(M/(M-V)),
\]
with zero contributions at the endpoints.

\begin{lemma}[Terminal credit]
\label{lem:terminal-credit}
Let $V(r)$ be the selected cell's mass and $D=R/\varepsilon$.
For any pre-choice subset $\mathcal Z$ of terminal active edges,
\[
 D\,\E\sum_{e\in\mathcal Z}a_e\mathbf1_{\{e\ {\rm cut}\}}
 \leq\frac{2e}{c_0\varepsilon}\E H_M(V(r)),
 \qquad c_0=(4-e)/3>0.
\]
\end{lemma}
\begin{proof}
Write $x=\Lambda R$ and $z=m/M$.
Eligibility gives
\[
 \beta\leq\frac{z^2}{4-z^2}\leq z^2/3\leq1/3.
\]
A terminal edge's first entry threshold lies above $R/2$ and within
$1/\Lambda$ of $R$.  The boundary identity therefore gives
\[
 P_{{\rm cut},e}\leq e\kappa_e\beta
                         \leq4e\ell_e\Lambda\beta.
\]
Both endpoints remain outside with probability at least
$1-\beta(e-1)\geq c_0$.
The endpoint budgets hold pointwise:
\[
 2\sum_{e\in\mathcal Z}b_e
     \mathbf1_{\{\text{both endpoints outside}\}}
                \leq\mu(Y\setminus X_p(r)).
\]
Taking expectations yields
$\sum_{e\in\mathcal Z}b_e\leq
\E\mu(Y\setminus X_p(r))/(2c_0)$.
The left side in the lemma is consequently at most
\[
 \frac{2e}{c_0\varepsilon}\,x\beta\,
                            \E\mu(Y\setminus X_p(r)).
\]
The function $x/(e^{x/2}-1)$ decreases for $x>0$.
Using $x\geq4\log(2/z)$ and
$\max_{0<z\leq1}z\log(2/z)=2/e$ gives
\[
 x\beta\leq\frac{4z^2\log(2/z)}{4-z^2}
 \leq\frac43z^2\log(2/z)\leq\frac8{3e}z<z.
\]
Every allowed cell contains its core, so $V\geq m$.
Moreover $\mu(Y\setminus X_p)\leq M-V$ and
$-\log(1-y)\geq y$ for $0\leq y<1$.  Hence
\[
 \frac mM\mu(Y\setminus X_p)
 \leq\frac VM(M-V)
 \leq(M-V)\log\frac M{M-V}\leq H_M(V).
\]
Substitution proves the claim.
\end{proof}

\begin{lemma}[Rational entropy credits]
\label{lem:rational-entropy}
For rational $0<V<M$, put
\[
 s=\min\{V,M-V\},\qquad k=\lceil\log_2(M/s)\rceil,
 \qquad U_M(V)=s(1+k).
\]
Then
\[
 H_M(V)\leq U_M(V)\leq\frac3{\log2}H_M(V)<6H_M(V).
\]
Over all actual source cells,
\[
 \sum_{X_p}U_{\mu(X)}(\mu(X_p))\leq12H_0.
\]
\end{lemma}
\begin{proof}
By symmetry take $V=s\leq M/2$.
The inequality $-\log(1-y)\leq y/(1-y)$ implies
\[
 s\log(M/s)\leq H_M(s)\leq s(\log(M/s)+1)\leq s(k+1).
\]
Since $k\leq\log(M/s)/\log2+1$ and $\log(M/s)\geq\log2$,
one also has $U_M(s)\leq(3/\log2)s\log(M/s)$.

For any partition into masses $v_i$,
$\sum_i H_M(v_i)\leq2\sum_i v_i\log(M/v_i)$.
To see this, normalize $p_i=v_i/M$ and let
$f(p)=-p\log p+(1-p)\log(1-p)$.
On $[0,1/2]$ the function is nonnegative and concave, with $f(0)=0$,
so $f(a)+f(b)\geq f(a+b)$ when $a+b\leq1/2$.
If every $p_i\leq1/2$, then $\sum_i f(p_i)\geq0$.
Otherwise the unique $p_1>1/2$ has complement
$q'=\sum_{i>1}p_i<1/2$, and
$\sum_i f(p_i)\geq f(p_1)+f(q')=0$.
This is the required binary-entropy inequality.

Include each retained protected path as a terminal part of its parent's
partition.  The complete partition entropies telescope to the entropy
of the final parts, which is at most $H_0$.
The omitted part credits are nonnegative; multiplication by the factor
six proves the total budget.
\end{proof}

\subsection{Selecting a radius}

For a radius in the open support interval away from every vertex entry
threshold, compute
\begin{align*}
 C(r)&=W\sum_{e\ {\rm active}}a_et_X(z_e)
                        \mathbf1_{\{\tau_e<r<\omega_e\}},\\
 Q(r)&=8W\Lambda\sum_{e\in\mathcal M}b_et_X(z_e)I_e
                        \mathbf1_{\{\tau_e<r\}},\\
 F&=4W\Lambda\frac{m_p^2}{3M^2}
                        \sum_{e\in\mathcal F}b_et_X(z_e)I_e,\\
 U(r)&=U_M\bigl(\mu\{v\in Y:D_p(v)<r\}\bigr).
\end{align*}
The cell is nonempty and excludes the nonempty protected path, so its
mass lies strictly between zero and $M$.

\begin{lemma}[Finite radius rule]
\label{lem:finite-radius}
Among the open intervals between consecutive distinct values of
\[
 \{R/2,R\}\cup\{D_p(v):v\in Y,\ R/2<D_p(v)<R\},
\]
there is one whose midpoint satisfies
\begin{equation}
\label{eq:local-radius-rule}
 C(r)\leq Q(r)+\frac{18\chi W}{\varepsilon}U(r)+F.
\end{equation}
Selecting the first such midpoint in increasing order uses only exact
rational arithmetic and comparisons with powers of two.
\end{lemma}
\begin{proof}
Use the auxiliary exponential density for this state only.
For far edges retain both terms of Lemma~\ref{lem:local-boundary};
for near nonterminal edges use its factor-two absorption.
After multiplication by the pre-choice bound
\eqref{eq:dyad-predraw-cost}, all main terms are bounded by
$\E Q(r)$.  The far remainder is at most $F$, since
$\beta\leq m_p^2/(3M^2)$.

For near terminal edges, $t_X(z_e)<\chi R/\varepsilon$.
Their entire expected cut cost is therefore at most
\[
 \frac{2e\chi W}{c_0\varepsilon}\E H_M(V(r))
 <\frac{18\chi W}{\varepsilon}\E U(r),
\]
by Lemma~\ref{lem:terminal-credit}, using $e<3$ and $c_0>1/3$.
Consequently
$\E C(r)\leq\E Q(r)+(18\chi W/\varepsilon)\E U(r)+F$.
Each function is constant on every listed open interval, and every
nonempty interval has positive auxiliary probability.
Thus at least one interval satisfies \eqref{eq:local-radius-rule}.

The portal is also constant on each interval: on the source path,
$D_p(v)=h(p)-h(v)$, so all portal transitions occur at listed vertex
thresholds.  The midpoint realizes the same cell and portal.
Neither an auxiliary probability nor a conditional expectation of
future cost is evaluated by the algorithm.
\end{proof}

\paragraph{An exact sweep for the radius.}
The following helper implements~\eqref{eq:local-radius-rule} without
sampling.  It receives the parent data and residual distance tables
already computed by $\textsc{Build}$.  In particular, $M=\mu(V(X))$
is the fixed parent mass, not the residual mass.  All events at the
same threshold are processed together; tested radii are strictly
between consecutive thresholds.

\begin{paperalgorithm}{\textsc{ChooseRadius}}
 {alg:choose-radius}
\textbf{Input:} The current parent and residual data
$(X,Y,P,S,h,t,R,\mu)$, threshold $\Lambda$, eligible source $p$,
its core mass $m_p$, current distance table $d_{Y,w}$, and active
residual edges $\mathcal A$.  The fixed weights $a_e,b_e$ are shared.

\textbf{Output:} A rational radius $r\in(R_p/2,R_p)$ satisfying
\eqref{eq:local-radius-rule}, and its cell $Z$.
\begin{enumerate}[label=\arabic*.,leftmargin=2em,itemsep=2pt,topsep=2pt]
\item Put $R=R_p$ and $M=\mu(V(X))$, and let $Q_p$ be the
root-to-$p$ path of $S$.
For every $v\in Y$ compute
\[
 D_p(v)=h(p)-2h(v)+\min_{q\in Q_p}
                 \{h(q)+2d_{Y,w}(q,v)\}.
\]
\item For each $e=uv\in\mathcal A$, compute
$\tau_e=\min\{D_p(u),D_p(v)\}$,
$\omega_e=\max\{D_p(u),D_p(v)\}$, and a fixed endpoint
$z_e$ attaining $\tau_e$.  Set
\[
 I_e=\mathbf1_{\{\max\{\tau_e,R/2\}<\min\{\omega_e,R\}\}}.
\]
Form the fixed sets
\begin{align*}
 \mathcal F&=\{e\in\mathcal A:I_e=1,\ R\leq R_{z_e}/\chi\},\\
 \mathcal M&=\mathcal F\ \cup\
   \{e\in\mathcal A:I_e=1,\ R>R_{z_e}/\chi,
                                  \ \tau_e\leq R-1/\Lambda\}.
\end{align*}
Precompute
\[
 c_e=W a_e t(z_e),\qquad
 q_e=8W\Lambda b_e t(z_e)\mathbf1_{\{e\in\mathcal M\}},
 \qquad
 F=\frac{4W\Lambda m_p^2}{3M^2}
                       \sum_{e\in\mathcal F}b_e t(z_e).
\]
\item Sort the distinct values in
$\{R/2,R\}\cup\{D_p(v):R/2<D_p(v)<R\}$ as
$\beta_0<\cdots<\beta_s$.  Record each vertex's threshold group.
Using these indices, bucket the vertex mass increments and the
active-edge entry and exit events at interior thresholds in one pass;
the update sums below are the resulting bucket totals.
For the first midpoint
$r_0=(\beta_0+\beta_1)/2$, initialize
\[
 Z=\{v:D_p(v)<r_0\},\quad V=\mu(Z),\quad
 C=\sum_{e:\,\tau_e<r_0<\omega_e}c_e,\quad
 Q=\sum_{e:\,\tau_e<r_0}q_e.
\]
\item For $i=0,\ldots,s-1$, in increasing order, set
$r=(\beta_i+\beta_{i+1})/2$ and compute
\[
 u=\min\{V,M-V\},\qquad
 k=\min\{j\in\mathbb Z_{\geq0}:2^j u\geq M\},
 \qquad U=u(1+k).
\]
If $C\leq Q+(18\chi W/\varepsilon)U+F$, return $(r,Z)$.
Otherwise, if another interval remains, process all events at
$d=\beta_{i+1}$ together:
\begin{align*}
 Z&\leftarrow Z\cup\{v:D_p(v)=d\},
 &V&\leftarrow V+\sum_{v:\,D_p(v)=d}\mu(v),\\
 C&\leftarrow C+\sum_{e:\,\tau_e=d}c_e
                    -\sum_{e:\,\omega_e=d}c_e,
 &Q&\leftarrow Q+\sum_{e:\,\tau_e=d}q_e.
\end{align*}
\end{enumerate}
\end{paperalgorithm}

Lemma~\ref{lem:finite-radius} guarantees that the helper returns.
The cell always has $0<V<M$, so the integer defining $U$ exists.
An edge with equal entry thresholds contributes cancelling events to
$C$, and has $q_e=0$.  Each vertex enters once and each active edge
has at most two events.  Consequently, once the gauges have been
computed, sorting and sweeping use
$O(|E(Y)|+|Y|\log|Y|)$ arithmetic operations.
Only the actual entry thresholds are sorted; no rational grid is
enumerated.  The portal chosen by $\textsc{Build}$ is constant on
the returned interval.

\begin{theorem}[Deterministic weighted spanning-tree bound]
\label{thm:dyad-weighted}
For positive physical lengths and nonnegative edge demands, the source
rule and finite radius rule return a spanning tree satisfying
\[
 C_a(T)\leq K_0B+K_1H_0
              \leq C_{\rm det}B\log n,
\]
where
\[
 K_1=8W(C_M+C_F)+4WC_T+\frac{216\chi W}{\varepsilon},\qquad
 C_{\rm det}=\frac{K_0}{\log2}+3K_1.
\]
The constant depends only on the fixed admissible parameters.
For rational input, the algorithm runs in deterministic polynomial
time in the standard Turing model.
\end{theorem}
\begin{proof}
Every edge is paid at its first small-scale parent or its first
separation.  These events exhaust its finite history, because the final
retained pieces are protected paths, on which endpoint scales are zero.
The first type contributes at most $K_0B$, and the second contributes
at most $\sum C(r)$.

At a selected radius, every main term in $Q(r)$ is assigned to an
endpoint $z_e$ actually captured by that source.
A far term is covered by $Q_F$.  For a near term, the stopping rule
implies
\[
 R_p>R_{z_e}/\chi>16\ell_e.
\]
Boundary overlap gives
$D_p(z_e)=\tau_e>R_p/2-4\ell_e>R_p/4$.
Thus the capture is flagged and covered by $Q_M$.
At any capture of a vertex, all assigned incident baseline weights sum
to at most its fixed $\mu$-mass.  Hence
\[
 \sum Q(r)\leq8W(Q_M+Q_F)\leq8W(C_M+C_F)H_0.
\]
Corollary~\ref{cor:dyad-rational-tail} gives
$\sum F\leq4WC_TH_0$, including missed far opportunities, and
Lemma~\ref{lem:rational-entropy} gives $\sum U(r)\leq12H_0$.
Summing \eqref{eq:local-radius-rule} proves the first bound.
All three credit budgets hold for every admissible trajectory.
Finally $H_0\leq3B\log n$ and $n\geq2$ imply the stated value of
$C_{\rm det}$.  Section~\ref{sec:weighted-running-time} proves the
running-time bound.
\end{proof}
\section{Running time}
\label{sec:weighted-running-time}

\begin{theorem}[Polynomial construction with exact dyadic eligibility]
\label{thm:weighted-running-time}
Let the positive physical lengths and nonnegative edge demands be
rational, and let $\mathsf L$ be their total binary input length,
including the graph encoding.  Use fixed rational parameters satisfying
\eqref{eq:admissible-parameters}, the fixed measure
$\mu(v)=\sum_{e\ni v}a_e\ell_e+B/n$, and the exact rules
\[
 K_v(Y)=\left\lceil\log_2\frac{2\mu(X)}{m_v(Y)}\right\rceil,
 \qquad \lambda_v(Y)=\frac{4K_v(Y)}{R_v},
 \qquad R_v=\varepsilon d_{X,\ell}(v,P_X).
\]
Here $B=\sum_ea_e\ell_e$; if $B=0$, any spanning tree suffices.
By Lemma~\ref{lem:finite-radius}, a suitable interval
midpoint can be found using only exact rational arithmetic and
integer comparisons.  The whole weighted-tree construction can be
implemented in
\[
 O(n^4+nm)
\]
arithmetic operations on numbers with
$O(\mathsf L+n+\log(2m))$ bits.  A conservative running-time
bound is therefore
\[
 O\!\left((n^4+nm)
        [\mathsf L+n+\log(2m)]^3\right).
\]
The bound is not optimized.  Numerical aspect ratios enter only
through their binary encodings.
\end{theorem}

\begin{proof}
Let $D$ be a common denominator of the original physical lengths,
obtained by multiplying their denominators.  Its bit length is
$O(\mathsf L)$.  Every current length is either an original physical
length or half that length: the construction never repeatedly halves
an inherited edge.  Thus current shortest-path distances and heights
are on the grid $1/(2D)$, and physical distances are on the grid
$1/D$.  Every shortest path is simple and has at most $n-1$ edges,
so their numerators have $O(\mathsf L+\log n)$ bits.

Write the fixed constants as $\varepsilon=a/b$ and
$\delta=c/d$ with positive integers $a,b,c,d$.
Then $R_v$ is on the grid $1/(bD)$, a physical core radius is on
the grid $1/(bdD)$, and vertex-entry gauges are on the grid
$1/(2D)$.  The endpoints $R_p/2,R_p$ and all entry gauges in
the allowed interval are therefore on the grid $1/(2bD)$.
Their interval midpoints can be taken on the grid $1/(4bD)$.
The algorithm enumerates only the at most $n+1$ intervals defined by
the actual gauges; it never enumerates the grid.

The fixed baseline weights $a_e\ell_e$ and all fixed vertex masses
have a common denominator of $O(\mathsf L+\log n)$ bits.  Every
subset mass has the same encoding bound.  Since
\[
 \mu(v)\geq B/n,\qquad \mu(X)\leq3B,
\]
every candidate satisfies
\[
 2\leq \frac{2\mu(X)}{m_v(Y)}\leq6n,
 \qquad 1\leq K_v(Y)\leq\lceil\log_2(6n)\rceil.
\]
The ceiling of the base-two logarithm of a positive rational $u/v$
is found exactly: use the difference of the bit lengths of the
positive integers $u,v$ to obtain one of the two possible integer
values, and resolve the choice by comparing $u$ with the appropriate
power-of-two multiple of $v$.  This uses integer shifts and one
comparison, and evaluates no logarithm.

Let $\ell_{\min},\ell_{\max}$ denote the smallest and largest
physical lengths.  For all personal rates, in every recursive
instance,
\[
 L_*:=\frac{4}{\varepsilon(n-1)\ell_{\max}}
 \leq\lambda_v(Y)
 \leq\frac{4\lceil\log_2(6n)\rceil}
              {\varepsilon\ell_{\min}}=:U_*.
\]
The initial threshold is one of the root's rational personal rates.
All later thresholds are dyadic multiples of it.  An internal
doubling can occur only below $U_*$, and its resulting value is less
than $2U_*$.  There are at most $n-1$ successive inheritances on a
branch, so every threshold is at most $2^{n+1}U_*$.  Its exponent
relative to the initial threshold is
\[
 O\bigl(n+\log(U_*/L_*)\bigr)
 =O(n+\mathsf L+\log n).
\]
Thus thresholds have $O(\mathsf L+n+\log n)$-bit encodings.
When a residual has no eligible source, all consecutive doublings
can be performed at once: determine the smallest dyadic multiple of
the current threshold at least the minimum personal rate.  The same
bit-length comparison computes this integer exponent exactly.
This produces precisely the threshold that consecutive doublings
would have produced.

Every selected source becomes part of its child's protected path.
That path is excluded from all subsequent source cells.  Hence a
vertex is selected as a source at most once; there are at most
$n-1$ source selections and at most $n$ recursive instances.

At each removal, recompute all-pairs physical distances and, if
desired, all-pairs current distances by Floyd--Warshall.  This uses
$O(n^3+m)$ arithmetic operations per residual and provides all
core masses in another $O(n^2)$ operations.  All exact dyadic rates
and the next source can then be determined in $O(n)$ operations.
One can instead compute the chosen source's gauge by a multi-source
shortest-path computation; the conservative all-pairs bound already
suffices.

Sort the at most $n$ entry gauges and sweep the open intervals.
Maintain the cell mass, the cut-cost sum, and the capture-credit sum;
each vertex enters once and each active edge has at most two entry
events.  The far credit is constant during this sweep.  The rational
binary-entropy majorant is computed from the current cell mass by
the same exact power-of-two comparison.  The sweep costs
$O(m+n\log n)$ arithmetic operations.  Source-path portal changes
are among these entry thresholds, so a selected midpoint determines
both the cell and its portal.

All local cost and credit expressions use subset masses, the fixed
baseline weights, physical distances, and the one current threshold.
Their encoding lengths are $O(\mathsf L+n+\log(2m))$.
For example, the far credit's extra denominator is the square of
the current parent mass; this contributes only
$O(\mathsf L+\log n)$ bits.  Parent-dependent credits are not
accumulated by the algorithm, so their denominators do not multiply
along a recursive chain.  Summing the operation counts over at most
$n-1$ removals proves the displayed arithmetic bound.  Exact
rational arithmetic with a conservative cubic time per arithmetic operation
gives the stated running-time bound.
\end{proof}

\paragraph{Composition with the rational master.}
Let $\mathsf L_0$ be the encoding length of the fixed metric or
capacity instance, and fix rational $0<\vartheta\leq1$.
The master in Theorem~\ref{thm:refined-rational-master}, applied in
dimension $m$, uses
\[
 N_{\rm master}=O\bigl(m\log(2m)/\vartheta^2\bigr)
\]
oracle calls.  Its per-coordinate query encoding is uniformly
\[
 Q=O\!\left(
   \frac{\log(2m)}{\vartheta^2}
       [1+\operatorname{bits}(\vartheta)]+\log(2m)
   \right).
\]
For distances the oracle demands are $p_e/\ell_e$; for capacities
the auxiliary lengths are $p_e/c_e$ and the demands are $c_e$.
In either case the weighted-oracle input has total encoding length
$O(\mathsf L_0+mQ)$.  In the capacity case every auxiliary length
is strictly positive because every master query coordinate is
strictly positive.

Consequently the total time spent in weighted-tree oracle calls is at most
\[
 O\!\left(
 N_{\rm master}(n^4+nm)
 [\mathsf L_0+mQ+n+\log(2m)]^3
 \right).
\]
Add the polynomial master cost proved in
Theorem~\ref{thm:refined-rational-master}, and the optional exact
support-reduction cost.  Each returned tree's fixed stretch or
congestion vector can be computed in $O(m\log n+n)$ arithmetic
operations using tree depths, lowest common ancestors, and subtree
sums; its entries have $O(\mathsf L_0+\log(2nm))$ bits,
independently of the changing oracle query.
This yields an explicit deterministic polynomial-time construction
for fixed $\vartheta$, and a polynomial dependence on
$1/\vartheta$ when accuracy is part of the input.  The displayed bound establishes polynomial running time; improving
it to near-linear time would require additional algorithmic work.
\section{Constructing an explicit distribution}
\label{sec:refined-rational-master}

The response-oracle approach of~\cite{CCGGP,Racke,AF,PST,AHK} converts a
weighted tree oracle to an explicit distribution.  This conversion
can preserve its approximation factor up to any prescribed multiplicative
loss.  The number of oracle calls need not depend on the numerical size
of the response bound, or on the aspect ratio of the graph data.  The
normalization of the stopping horizon is essential to this statement.

\begin{theorem}[Rational adaptive-step master]
\label{thm:refined-rational-master}
Let $d\geq 1$, let $\mathcal A\subseteq\mathbb Q_{\geq0}^{d}$ be a
finite family of nonnegative rational vectors, and let $\rho>0$ be
rational.  Suppose that for every strictly positive rational probability
vector $p$ an oracle returns $a\in\mathcal A$ satisfying
\[
 p\cdot a\leq\rho.
\]
For every rational $0<\vartheta\leq1$, a deterministic algorithm uses at
most
\[
 N\leq 1+\left\lceil
       \frac{20d\lceil\log_2(2d)\rceil}{\vartheta^2}
       \right\rceil
\]
oracle calls and returns rational probabilities $\lambda_t$ and oracle
outputs $a^{(t)}$ such that
\[
 \sum_t\lambda_t=1,
 \qquad
 \sum_t\lambda_t a^{(t)}_j\leq(1+\vartheta)\rho
 \quad(1\leq j\leq d).
\]
No exponential or logarithmic function is evaluated by the algorithm.
Its oracle queries have a uniform polynomial encoding bound in
$d$, $1/\vartheta$, and the encoding of $\vartheta$, independent of
the numerical magnitude and encoding length of $\rho$.

If the oracle runs in polynomial time in the fixed input and the query
encoding, and its returned vectors have polynomial encoding length in
these parameters, then the entire construction has polynomial running
time in the fixed input, $1/\vartheta$, and the encodings of
$\rho$ and $\vartheta$.  In particular, numerical aspect ratios impose
no pseudopolynomial dependence.
\end{theorem}

\begin{proof}
Put
\[
 k=\lceil\log_2(2d)\rceil,
 \qquad \xi=\vartheta/4,
 \qquad b=1+\xi,
 \qquad H=\frac{10k}{\vartheta^2\rho}.
\]
The integer $k$ is computed by comparisons with powers of two.  All
other quantities in this display are rational.  Initialize
$x=0\in\mathbb Q^d$ and $\tau=0$.  While $\tau<H$, form
\[
 w_j=b^{\lceil x_j\rceil},\qquad
 p_j=\frac{w_j}{\sum_iw_i},
\]
and obtain a response $a$ from the oracle.  Write $L=\max_j a_j$ and
choose
\[
 \Delta=
 \begin{cases}
  H-\tau,&L=0,\\
  \min\{L^{-1},H-\tau\},&L>0.
 \end{cases}
\]
Record the pair $(\Delta/H,a)$ and update
$x\leftarrow x+\Delta a$ and $\tau\leftarrow\tau+\Delta$.
Every coordinate increment lies in $[0,1]$.

For analysis only, define $\Phi(x)=\sum_j b^{x_j}$.
Convexity of $b^z$ on $[0,1]$ gives
$b^z\leq1+(b-1)z=1+\xi z$.  Moreover,
$b^{x_j}\leq w_j\leq b\,b^{x_j}$, so
\[
 \sum_j b^{x_j}a_j
 \leq\sum_jw_ja_j
 =\left(\sum_jw_j\right)(p\cdot a)
 \leq b\rho\Phi(x).
\]
It follows that
\[
 \Phi(x+\Delta a)
 \leq \Phi(x)+\xi\Delta\sum_jb^{x_j}a_j
 \leq \Phi(x)(1+\xi b\rho\Delta)
 \leq \Phi(x)\exp(\xi b\rho\Delta).
\]
Induction therefore gives, at every reached state,
\[
 \Phi(x)\leq d\exp(\xi b\rho\tau),\qquad
 x_j\leq \frac{\ln d}{\ln b}
          +\frac{\xi b\rho\tau}{\ln b}.
\]

Here are explicit elementary estimates for the two coefficients.
Since $0<\xi\leq1/4$,
\[
 \ln(1+\xi)\geq\frac{\xi}{1+\xi}
 \geq\frac{4\xi}{5}=\frac{\vartheta}{5}.
\]
Also $\ln(1+\xi)\geq\xi-\xi^2/2$, and hence
\[
 \frac{\xi b}{\ln b}
 \leq\frac{1+\xi}{1-\xi/2}
 \leq1+2\xi=1+\frac{\vartheta}{2}.
\]
The penultimate inequality follows by multiplying its positive
denominator; the difference of the resulting right and left sides is
$\xi(1/2-\xi)\geq0$.
Since $\ln d\leq k$ and $\tau\leq H$, all coordinates satisfy
\begin{align*}
 x_j
 &\leq \frac{5k}{\vartheta}
       +\left(1+\frac{\vartheta}{2}\right)
          \frac{10k}{\vartheta^2}\\
 &=\frac{10k}{\vartheta^2}+\frac{10k}{\vartheta}
 \leq\frac{20k}{\vartheta^2}=:\Gamma.
\end{align*}

Every nonfinal iteration has $L>0$ and $\Delta=1/L$, and raises at
least one coordinate of $x$ by exactly one.  Thus it raises
$\sum_jx_j$ by at least one.  This sum never exceeds $d\Gamma$;
there are at most $\lfloor d\Gamma\rfloor$ nonfinal iterations.
The algorithm consequently terminates within the asserted call bound.
At termination the recorded coefficients sum to one, and their mean
vector is $x/H$.  Finally,
\[
 \frac{x_j}{H}
 \leq\frac{\ln d}{H\ln b}
       +\frac{\xi b}{\ln b}\rho
 \leq\frac{\vartheta\rho}{2}
       +\left(1+\frac{\vartheta}{2}\right)\rho
 =(1+\vartheta)\rho.
\]
The remaining assertions follow from the running-time analysis below.
\end{proof}

\paragraph{Uniform query encodings.}
Write $b=P/Q$ in lowest terms and let $B_\vartheta$ bound the bit
lengths of $P$ and $Q$.  At every iteration set
$e_j=\lceil x_j\rceil$ and $E=\max_j e_j\leq\lceil\Gamma\rceil$.
The query can be computed using the positive integers
\[
 W_j=P^{e_j}Q^{E-e_j},\qquad p_j=W_j/\sum_iW_i.
\]
Each coordinate of $p$ has bit length
\[
 O\bigl(\Gamma B_\vartheta+\log(2d)\bigr).
\]
This bound contains neither a previous query size nor a previous
oracle output size.  In particular, the polynomial-time oracle
hypothesis is not being iterated through a potentially growing sequence
of polynomial bounds.  Powers of $P$ and $Q$ through
$\lceil\Gamma\rceil$ can be precomputed once, after which each query
uses $O(d)$ integer arithmetic operations.

\paragraph{Accumulated rational encodings and an explicit master bound.}
Let $B$ be a uniform bound on the numerator and denominator lengths
of the returned vector entries, and let $B_H$ be the encoding length
of $H$.  Every nonfinal step size is the reciprocal of one returned
entry and has $O(B)$ bits.  The final step is $H$ minus the preceding
sum of step sizes.  A common-denominator representation consequently
bounds the encoding lengths of all accumulated times, load coordinates,
and output probabilities by
\[
 O\bigl(NB+B_H+\log(2N)\bigr).
\]
For example, before the final iteration each load coordinate is a sum
of at most $N$ products of two $B$-bit rationals.  The last iteration
adds one product involving the final step size; its denominator adds
at most the same total bound.  The normalization by $H$ adds $O(B_H)$
bits.  The magnitudes of the accumulated load coordinates are bounded
by $\Gamma$ as proved above.

Thus, putting
\[
 S=O\bigl(NB+B_H+\Gamma B_\vartheta+\log(2dN)\bigr),
\]
the master needs $O(Nd+\Gamma)$ arithmetic operations on numbers of
at most $S$ bits.  With conservative elementary integer arithmetic,
its additional running time is
$O((Nd+\Gamma)S^3)$, besides the $N$ oracle invocations and the
computation of returned payoff vectors.  This is an explicit
polynomial bound; faster arithmetic can improve it.  The dependence
on $\rho$ is through $B_H=O(\operatorname{bits}(\rho)+
\operatorname{bits}(\vartheta)+\log(2k))$, not its numerical value.
The same uniform query bound shows that a polynomial-time oracle has
a uniform polynomial response bound $B$ even if this bound is not
specified separately in advance.

\paragraph{Why the horizon may be smaller than one.}
The proof only requires $H>0$; there is no lower bound on its numerical
value.  Imposing $H\geq1$ would unnecessarily introduce a dependence
on the magnitude of $\rho$ into the maximum exponent and the call
count.  Choosing $H$ proportional to $1/\rho$ cancels that dependence
without changing the normalized output distribution.

\begin{corollary}[Support at most $d$ without increasing the mean]
\label{cor:refined-support-d}
Given the list returned by Theorem~\ref{thm:refined-rational-master},
an elementary procedure using only exact rational linear algebra finds
a distribution supported on at most $d$ of its vectors whose mean is
no larger in any coordinate.  Stopping one step earlier preserves the
entire mean vector and uses at most $d+1$ vectors.
Both procedures have polynomial running time; no linear program or
additional oracle call is needed.
\end{corollary}

\begin{proof}
First, while more than $d+1$ coefficients are
positive, find an integer relation
$\sum_t\alpha_t(1,a^{(t)})=0$ on any $d+2$ supported columns.
It has both signs.  With
$\theta=\min_{\alpha_t>0}\lambda_t/\alpha_t$, replace
$\lambda_t$ by $\lambda_t-\theta\alpha_t$.  This keeps all
probabilities nonnegative, preserves their sum and their full mean,
and removes a supported vector.  The columns never change, so rational
Gaussian elimination gives integer relations of uniformly polynomial
bit length.  If $\lambda_t=N_t/D$ and $i$ attains the minimum, the
shared-denominator update
\[
 N'_t=N_t\alpha_i-N_i\alpha_t,\qquad D'=D\alpha_i
\]
adds at most $\operatorname{bits}(\alpha_i)$ to the denominator
length.  The nonnegative new numerators sum to $D'$, so their lengths
are bounded by the same quantity.  There are at most $N$ elimination
steps, proving the stated bit bound for exact mean preservation.

If at most $d$ coefficients remain, we are done.  Otherwise there are
exactly $d+1$ positive coefficients.  If their augmented columns
$(1,a^{(t)})$ are dependent, apply the same elimination once more.
If they are independent, solve the nonsingular rational system
\[
 \sum_t h_t=0,\qquad
 \sum_t h_t a^{(t)}=-\mathbf1_d.
\]
The vector $h$ is nonzero and has both signs because its entries sum
to zero.  Put
\[
 t_* = \min_{h_t<0}\frac{\lambda_t}{-h_t}>0,
 \qquad \lambda'_t=\lambda_t+t_*h_t.
\]
All new probabilities are nonnegative and sum to one, at least one
becomes zero, and the mean changes by $-t_*\mathbf1_d$.
Thus every coordinate decreases and at most $d$ vectors remain.
Because the probabilities remain a convex combination of nonnegative
vectors, this step cannot make any mean coordinate negative.

For completeness, the final system has a polynomial-bit rational
solution by determinant bounds on its unchanged input columns.
Clearing its denominators gives an integer vector $\alpha$ and a
positive integer $q$ such that
$\sum_t\alpha_t=0$ and
$\sum_t\alpha_ta^{(t)}=-q\mathbf1_d$.
If $i$ minimizes $\lambda_i/(-\alpha_i)$ over negative
$\alpha_i$, the shared-denominator update for this last move is
\[
 N'_t=N_t(-\alpha_i)+N_i\alpha_t,
 \qquad D'=D(-\alpha_i).
\]
Its numerators are nonnegative and sum to $D'$.  It therefore has
the same polynomial bit-growth bound as an ordinary elimination.
\end{proof}

\paragraph{Constructing the final probabilities.}
Algorithm~\ref{alg:rational-master} gives the adaptive mixture and
support reduction.  Each oracle response contains a tree and its
payoff vector; the final loop changes only their probabilities.

\begin{paperalgorithm}{Rational mixture and support reduction}
 {alg:rational-master}
\textbf{Input:} A dimension $d\geq1$, rational parameters $\rho>0$ and
$0<\vartheta\leq1$, and an exact oracle which, for each strictly
positive rational probability vector $p$, returns $(T,v)$ with
$v\in\mathbb Q_{\geq0}^d$ and $p\cdot v\leq\rho$.

\textbf{Output:} Trees $T_t$ and positive rational probabilities
$q_t$ with support at most $d$, $\sum_tq_t=1$, and
$\sum_tq_tv^{(t)}\leq(1+\vartheta)\rho\mathbf1$ coordinatewise.
\begin{enumerate}[label=\arabic*.,leftmargin=2em,itemsep=2pt,topsep=2pt]
\item Compute $k=\min\{j\in\mathbb Z_{\geq0}:2^j\geq2d\}$,
$b=1+\vartheta/4$, and $H=10k/(\vartheta^2\rho)$.
Write $b=P_b/Q_b$ in lowest terms, set
$J=\lceil20k/\vartheta^2\rceil$, and precompute the integer powers
$P_b^s,Q_b^s$ for $0\leq s\leq J$.
Initialize $x=0\in\mathbb Q^d$, $\tau=0$, and an empty response list.
\item \textbf{While $\tau<H$:} compute
\[
 e_j=\lceil x_j\rceil,\qquad
 w_j=P_b^{e_j}Q_b^{J-e_j},\qquad
 p_j=w_j/\sum_iw_i,
\]
and obtain $(T,v)$ from the oracle.  Put $L=\max_jv_j$ and set
$\Delta=H-\tau$ if $L=0$, or
$\Delta=\min\{1/L,H-\tau\}$ otherwise.
Append $(\Delta/H,T,v)$ to the list and update
$x\leftarrow x+\Delta v$ and $\tau\leftarrow\tau+\Delta$.
\item Coalesce identical pairs $(T,v)$, adding their probabilities.
While more than $d$ positive coefficients remain, let $A$ have
the supported augmented columns $(1,v^{(t)})$.
\begin{enumerate}[label=(\alph*),leftmargin=2em,itemsep=2pt,topsep=2pt]
\item If the columns are dependent, find a nonzero integer vector
$\alpha$ with $A\alpha=0$, supported on at most $d+2$ columns.
Set $s=\min_{\alpha_t>0}q_t/\alpha_t$ and replace
$q_t$ by $q_t-s\alpha_t$.
\item Otherwise there are exactly $d+1$ columns.  Solve
$Ah=(0,-\mathbf1_d)$, set
$s=\min_{h_t<0}q_t/(-h_t)$, and replace $q_t$ by $q_t+sh_t$.
\item Delete every zero coefficient and its response.
\end{enumerate}
\item Return the remaining pairs $(q_t,T_t)$.
\end{enumerate}
\end{paperalgorithm}

The integer weights in Step~2 give exactly the normalized prices
$p_j=b^{\lceil x_j\rceil}/\sum_i b^{\lceil x_i\rceil}$.
Theorem~\ref{thm:refined-rational-master} gives $e_j\leq J$;
the precomputed powers therefore give $O(d)$ arithmetic operations
per query, as in the uniform-encoding analysis.
For either graph application take $d=m$ and the response bound
$\rho=\rho_n$ from Section~\ref{sec:choice-of-parameters}.
For distances, the oracle calls
$\textsc{WeightedTree}(G,\ell,(p_e/\ell_e)_e)$ and returns the
stretch vector $v=s_T$ of~\eqref{eq:cap-distance-payoff}.
For capacities, it calls
$\textsc{WeightedTree}(G,(p_e/c_e)_e,c)$ and returns the
fundamental-cut congestion vector $v=g_T$
of~\eqref{eq:cap-capacity-payoff}.
These payoffs are evaluated against the fixed original length or
capacity vector, respectively.  Each weighted-tree call constructs
its own measure $\mu$, which remains fixed within that call.
The output probabilities define $q^{\rm dist}$ or $q^{\rm cap}$;
in the latter case assign every supported tree edge its fundamental-cut
capacity from~\eqref{eq:cap-fundamental-capacity}.
The call bound and polynomial running time follow from
Theorem~\ref{thm:refined-rational-master}; the last loop is precisely
the rational support reduction of Corollary~\ref{cor:refined-support-d}.

\section{Fundamental-cut capacities and routing}
\label{sec:capacity-extension}

Theorem~\ref{thm:dyad-weighted} supplies the weighted-tree hypothesis
used below.  We state the reduction for a generic response bound $\rho$.
The distance--capacity equivalence is the family-preserving framework
of Andersen--Feige~\cite{AF}, following R\"acke~\cite{Racke};
we reproduce the identities and routing argument to specify all quantifiers.

\subsection{The weighted hypothesis and the two payoff vectors}

Let $G=(V,E)$ be a finite connected undirected graph, with $n\geq2$
and $m=|E|\geq1$.  Loops may be deleted.  Parallel edges cause no
problem; a reduction to the simple-graph weighted hypothesis is given
below.  Write $\mathcal T(G)$ for the finite collection of its actual
spanning trees.  The weighted hypothesis with parameter $\rho\geq1$
is the following assertion:
\begin{equation}
\label{eq:cap-weighted-hypothesis}
 \text{for all }\ell_e>0,\ a_e\geq0,\quad
 \min_{T\in\mathcal T(G)}\sum_{e=uv}a_e d_{T,\ell}(u,v)
 \leq\rho\sum_e a_e\ell_e.
\end{equation}
Theorem~\ref{thm:dyad-weighted} gives a deterministic oracle
for this hypothesis with $\rho=\rho_n$.

For a tree $T$ and original edge $e=uv$, let $P_T(e)$ be the unique
$u$--$v$ path in $T$.  Set
\[
 M_T(e,f)=\mathbf1_{\{f\in P_T(e)\}},\qquad e,f\in E.
\]
For fixed original lengths $\ell>0$ the stretch payoff is
\begin{equation}
\label{eq:cap-distance-payoff}
 s_T(e)=\frac{d_{T,\ell}(e)}{\ell_e}
       =\frac{\sum_f M_T(e,f)\ell_f}{\ell_e}.
\end{equation}
For fixed original capacities $c>0$ the load and congestion payoffs
are
\begin{equation}
\label{eq:cap-capacity-payoff}
 L_T(f)=\sum_e c_e M_T(e,f),\qquad
 g_T(f)=\frac{L_T(f)}{c_f}.
\end{equation}
In particular $L_T(f)=0$ if $f\notin T$.  If $f\in T$ and $S_f$
is either component of $T-f$, then
\begin{equation}
\label{eq:cap-fundamental-capacity}
 L_T(f)=c_G(\delta_G(S_f)).
\end{equation}
We equip the spanning-tree topology with capacities
$c_T(f)=L_T(f)$ for $f\in T$.  The tree uses original vertices and
edges, while its capacities are the capacities of their fundamental
cuts; in general $c_T(f)$ differs from the original $c_f$.
The probabilistic congestion parameter is
$\max_f\mathbb E g_T(f)$, with the expectation taken before the
maximum.  A bound on this parameter need not bound
$\mathbb E\max_fg_T(f)$ for a single sampled tree.  The routing
guarantee below applies to the fixed fractional mixture of tree
paths.

\begin{lemma}[The weighted distance--capacity identity]
\label{lem:cap-weighted-identity}
For every strictly positive edge-price vector $w$ and every tree $T$,
the choices
\[
 \ell_f^{\,w}=\frac{w_f}{c_f},\qquad a_e=c_e
\]
satisfy
\begin{equation}
\label{eq:cap-weighted-identity}
 \sum_e a_e d_{T,\ell^{\,w}}(e)
 =\sum_f w_f g_T(f),\qquad
 \sum_e a_e\ell_e^{\,w}=\sum_e w_e.
\end{equation}
For fixed lengths $\ell$, the alternative choices
$a_e=w_e/\ell_e$ satisfy
\begin{equation}
\label{eq:cap-distance-oracle-identity}
 \sum_e a_e d_{T,\ell}(e)=\sum_e w_e s_T(e),\qquad
 \sum_e a_e\ell_e=\sum_e w_e.
\end{equation}
\end{lemma}

\begin{proof}
Exchange the two finite sums:
\[
 \sum_e c_e\sum_f M_T(e,f)\frac{w_f}{c_f}
 =\sum_f\frac{w_f}{c_f}\sum_e c_eM_T(e,f).
\]
The second identity in \eqref{eq:cap-weighted-identity} and both
identities in \eqref{eq:cap-distance-oracle-identity} are direct
substitutions.  No tree transformation, Steiner vertex, or change of
the admissible family occurs.
\end{proof}

\subsection{Exact existential bounds}

\begin{theorem}[Two distributions on the same family of spanning trees]
\label{thm:cap-existential-duality}
Assume \eqref{eq:cap-weighted-hypothesis}.  For every fixed $\ell>0$
there is a distribution $q^{\mathrm{dist}}$ on $\mathcal T(G)$ such
that
\[
 \mathbb E_{q^{\mathrm{dist}}}d_{T,\ell}(u,v)
 \leq\rho d_{G,\ell}(u,v)\qquad(u,v\in V).
\]
Every sampled tree also satisfies
$d_{T,\ell}(u,v)\geq d_{G,\ell}(u,v)$.
For every fixed $c>0$ there is a distribution
$q^{\mathrm{cap}}$ on the same family such that
\begin{equation}
\label{eq:cap-expected-edge-load}
 \mathbb E_{q^{\mathrm{cap}}}L_T(f)\leq\rho c_f
 \qquad(f\in E).
\end{equation}
The two distributions are allowed to differ.
\end{theorem}

\begin{proof}
For the distance assertion, let $p$ be any probability vector on
$E$.  Use $a_e=p_e/\ell_e$ in
\eqref{eq:cap-weighted-hypothesis}.  This gives
$\min_T p\cdot s_T\leq\rho$.  The finite minimax theorem produces
a distribution satisfying $\mathbb E s_T(e)\leq\rho$ for every
edge.  Summing the tree triangle inequality along a fixed shortest
graph path proves the all-pairs assertion.  Each tree path is a graph
path with the original lengths, which proves domination.

For capacities, first suppose $p_f>0$ for every edge and use
Lemma~\ref{lem:cap-weighted-identity} with $w=p$.  Then
$\min_Tp\cdot g_T\leq\rho$.  To allow zero prices, for any
$\eta>0$ use $w_f=p_f+\eta/m$.  Nonnegativity gives
\[
 \min_Tp\cdot g_T
 \leq\min_Tw\cdot g_T
 \leq\rho(1+\eta).
\]
Let $\eta\downarrow0$.  The left side is fixed, so it is at most
$\rho$.  Finite minimax now gives $\mathbb E g_T(f)\leq\rho$
for all $f$, which is \eqref{eq:cap-expected-edge-load}.
\end{proof}

\subsection{All-cut domination and the full routing statement}

\begin{proposition}[Every tree dominates all cuts]
\label{prop:cap-cut-domination}
For every spanning-tree $T$ equipped with
\eqref{eq:cap-fundamental-capacity}, and every $A\subseteq V$,
\begin{equation}
\label{eq:cap-all-cut-domination}
 c_G(\delta_G(A))\leq c_T(\delta_T(A)).
\end{equation}
Any distribution with
$\mathbb E L_T(f)\leq\alpha c_f$ for every original edge also
satisfies
\begin{equation}
\label{eq:cap-all-cut-expectation}
 \mathbb E c_T(\delta_T(A))
 \leq\alpha c_G(\delta_G(A))\qquad(A\subseteq V).
\end{equation}
\end{proposition}

\begin{proof}
By interchanging finite sums,
\[
 c_T(\delta_T(A))
 =\sum_e c_e\,|P_T(e)\cap\delta_T(A)|.
\]
If $e$ crosses $A$, its tree path crosses $A$ at least once.  This
proves \eqref{eq:cap-all-cut-domination}; paths for other edges may
contribute additional nonnegative terms.  For the upper bound,
\[
 c_T(\delta_T(A))
 =\sum_{f\in\delta_G(A)}L_T(f),
\]
where non-tree edges contribute zero.  Take expectations and apply
the individual edge-load bounds.
\end{proof}

\begin{corollary}[Cuts of prescribed cardinality]
\label{cor:cardinality-cuts}
Given the capacity distribution of Theorem~\ref{thm:main}, for every
$1\leq s\leq n-1$ one can find in polynomial time a set $S$ of
exactly $s$ vertices whose graph-cut capacity is at most $\alpha_n$
times the minimum possible capacity of such a set.  In particular,
for even $n$ this gives an $O(\log n)$ approximation for minimum
bisection with exact balance.
\end{corollary}
\begin{proof}
For each supported tree $T$, compute a minimum-capacity tree cut
$S_T$ with $|S_T|=s$.  Return the candidate with the smallest
original graph-cut capacity.  If $S^*$ is an optimal graph cut of
cardinality $s$, the expectation bound implies that some supported
tree satisfies
\[
 c_T(\delta_T(S^*))\leq\alpha_n c_G(\delta_G(S^*)).
\]
For that tree, domination and the definition of $S_T$ give
\[
 c_G(\delta_G(S_T))\leq c_T(\delta_T(S_T))
 \leq c_T(\delta_T(S^*))
 \leq\alpha_n c_G(\delta_G(S^*)).
\]
It remains to describe the tree computation.

Root $T$ arbitrarily.  The state $\mathrm{DP}_v(\sigma,j)$ is the
minimum cut capacity inside the subtree at $v$, when exactly $j$
of its vertices belong to the chosen set and $v$ has membership
$\sigma\in\{0,1\}$.  Process the children $w_1,\ldots,w_d$ of
$v$ in any fixed order.  Before processing them, set
$\mathrm{DP}_{v,0}(\sigma,j)=0$ if $j=\sigma$, and $+\infty$
otherwise.  Then use the knapsack recurrence
\[
 \mathrm{DP}_{v,i}(\sigma,j)=
 \min_{\substack{\tau\in\{0,1\}\\0\leq h\leq |V(T_{w_i})|}}
 \left\{
 \mathrm{DP}_{v,i-1}(\sigma,j-h)
 +\mathrm{DP}_{w_i}(\tau,h)
 +c_T(vw_i)\mathbf1_{\{\sigma\ne\tau\}}
 \right\},
\]
with invalid counts assigned $+\infty$, and set
$\mathrm{DP}_v=\mathrm{DP}_{v,d}$.  The optimum at the root $r$ is
\[
 \min_{\sigma\in\{0,1\}}\mathrm{DP}_r(\sigma,s).
\]
Storing a minimizing choice in each state recovers $S_T$.
The computation uses at most $O(n^3)$ rational arithmetic operations
per tree.  Evaluating its graph cut takes $O(m)$ more operations.
The rational capacities have polynomial encodings, and the support
has at most $m$ trees, so the complete procedure runs in polynomial
time.
\end{proof}

R\"acke~\cite{Racke} established the logarithmic approximation for
minimum bisection using hierarchical decomposition trees and a
dynamic program that counts terminals at the leaves.  The corollary
recovers that approximation from spanning trees; here the cardinality
state counts the original graph vertices directly.

\begin{proposition}[Canonical capacities are the smallest cut dominator]
\label{prop:cap-minimal-cut-capacities}
Fix an actual spanning-tree topology $T$. A nonnegative capacity
assignment $b_T$ on its edges dominates every cut of $G$ if and
only if
\[
 b_T(f)\geq c_G(\delta_G(S_f))\qquad(f\in T).
\]
Thus the fundamental-cut assignment $c_T$ is the coordinatewise
smallest capacity assignment on that topology with all-cut
domination.
\end{proposition}

\begin{proof}
For necessity, the fundamental cut $S_f$ crosses exactly one edge
of $T$, namely $f$. Its domination inequality is precisely the
displayed inequality. For sufficiency, the displayed inequalities
give $b_T(\delta_T(A))\geq c_T(\delta_T(A))$ for every cut $A$;
Proposition~\ref{prop:cap-cut-domination} supplies
$c_T(\delta_T(A))\geq c_G(\delta_G(A))$.
\end{proof}

For precision, a demand vector has one nonnegative coordinate
$D_{uv}$ for each unordered pair of distinct vertices.  Let
$\mathcal F(G,c)$ be the set of such vectors routable in $G$ by a
fractional multicommodity flow of congestion at most one.  Let
$\operatorname{opt}_{G,c}(D)$ be its minimum possible congestion.
For a cut $S$, write
\[
 D(S)=\sum_{\{u,v\}:|\{u,v\}\cap S|=1}D_{uv}.
\]
Every flow routing $D$ at congestion $\lambda$ satisfies the
necessary cut inequalities
$D(S)\leq\lambda c_G(\delta_G(S))$.

\begin{theorem}[Tree-flow mixtures and an oblivious routing operator]
\label{thm:cap-routing}
Suppose a distribution $q$ satisfies
$\mathbb E_q L_T(f)\leq\alpha c_f$ for all $f\in E$.
Then the following assertions hold.
\begin{enumerate}
\item For every tree $T$ and every demand vector $D$,
\[
 \operatorname{opt}_{T,c_T}(D)
 \leq\operatorname{opt}_{G,c}(D).
\]
In particular $\mathcal F(G,c)\subseteq\mathcal F(T,c_T)$.
\item For arbitrary demand vectors $D^T\in\mathcal F(T,c_T)$,
one for each supported tree, the averaged vector
$\sum_Tq_TD^T$ is routable in $G$ at congestion at most $\alpha$.
Consequently, with the sum denoting the weighted Minkowski sum,
\begin{equation}
\label{eq:cap-minkowski-sandwich}
 \mathcal F(G,c)
 \subseteq\sum_Tq_T\mathcal F(T,c_T)
 \subseteq\alpha\mathcal F(G,c).
\end{equation}
\item Route each pair $u,v$ by putting fraction $q_T$ of its demand
on $P_T(u,v)$.  This defines a fixed fractional oblivious routing
operator, independent of $D$, whose congestion for every $D$ is at
most $\alpha\operatorname{opt}_{G,c}(D)$.
\end{enumerate}
\end{theorem}

\begin{proof}
Unique path routing in a tree puts load exactly $D(S_f)$ on its
edge $f$.  If $D$ can be routed in $G$ at congestion $\lambda$,
the cut inequality for $S_f$ gives
\[
 D(S_f)\leq\lambda c_G(\delta_G(S_f))
          =\lambda c_T(f).
\]
Taking the minimum possible $\lambda$ proves the first assertion.

For the second assertion, route $D^T$ on the original paths of $T$
and scale that flow by $q_T$.  Its load on an original edge $f$ is
at most $q_TL_T(f)$.  Summing the flows gives the required averaged
demand vector and edge load at most
$\sum_Tq_TL_T(f)\leq\alpha c_f$.  Taking $D^T=D$ for all $T$
also proves the first inclusion in
\eqref{eq:cap-minkowski-sandwich}.

Finally, for a fixed $D$ and
$\lambda=\operatorname{opt}_{G,c}(D)$, the first assertion bounds
every tree's routing load by $\lambda c_T(f)$.  The second argument
then gives averaged edge load at most $\lambda\alpha c_f$.
All pair paths and mixture coefficients were fixed before $D$ was
specified, as required for oblivious routing.
\end{proof}

These are statements about one fixed distribution working for every
cut and demand vector.  The cut upper bound is an expectation for
each fixed cut.  It does not assert that one sampled tree
simultaneously has distortion $\alpha$ on all cuts.  A demand
feasible in a single supported tree need not be feasible in $G$ at
congestion $\alpha$; the weighted mixture statement in
Theorem~\ref{thm:cap-routing} is the relevant converse.

\begin{proof}[Proof of Theorem~\ref{thm:main}]
Apply Theorem~\ref{thm:dyad-weighted} with bound $\rho_n$.
For the distance instance, a positive master query $p$ uses
$a_e=p_e/\ell_e$ and returns the stretch vector $s_T$.
For the capacity instance use $\ell_e=p_e/c_e$ and $a_e=c_e$,
which returns the congestion vector $g_T$ by
Lemma~\ref{lem:cap-weighted-identity}.  Both queries have baseline one.
Theorem~\ref{thm:refined-rational-master} therefore constructs an
explicit rational distribution whose every coordinate is at most
$(1+\vartheta)\rho_n$.
Corollary~\ref{cor:refined-support-d} reduces its support to at most
$m$ without increasing any expected coordinate.

For distances, summing the tree triangle inequality along a fixed
shortest graph path gives the all-pairs bound.  Every tree path is an
original graph path, so metric domination is deterministic.
For capacities, Proposition~\ref{prop:cap-cut-domination} and
Theorem~\ref{thm:cap-routing} give the cut and routing conclusions
with the same factor.  All graph payoffs are sums of fixed input
rationals divided by a fixed positive input rational.
Section~\ref{sec:weighted-running-time} and the master's bit analysis
give the claimed running time, probabilities, and call bound.
For positive real input, approximate the lengths or capacities by
positive rationals and pass to a convergent subsequence of distributions
on the finite tree family.  There are only finitely many supports of
size at most $m$, so the limit retains this support bound.
All distance and load inequalities are continuous in the input.
\end{proof}

\paragraph{Computing all fundamental-cut capacities.}
Root the returned tree and preprocess lowest common ancestors.
For every graph edge $uv$, add $c_{uv}$ at $u$ and $v$, and subtract
$2c_{uv}$ at their lowest common ancestor.
Taking subtree sums makes the sum below each nonroot vertex equal
to the capacity of its parent edge's fundamental cut.
This computes all tree capacities in $O(m\log n+n)$ arithmetic
operations.
\paragraph{Parallel edges and zero capacities.}
For parallel edges with common endpoints, retain one edge of minimum
auxiliary length and replace their demands by their sum.  The
weighted tree cost is unchanged, while its baseline weakly
decreases.  Therefore a weighted theorem proved for simple graphs
extends to multigraphs with the same constant, and the chosen
representative is an original edge.
For nonnegative capacities, first form the connected components of
the positive-capacity subgraph.  Run the capacity construction
inside each such component and join those component trees by a
fixed spanning tree of zero-capacity edges in the quotient.  A
joining edge has fundamental-cut capacity zero.  A joint oracle
call applies the weighted oracle separately to every positive
component; their weighted costs add.  The rational master on
the coordinates of positive-capacity edges consequently produces
a polynomial-support distribution for the whole graph, without
taking an exponentially large product of component supports.
The all-zero-capacity case is immediate.  A one-vertex graph is
also immediate.

\section{Lower bounds}
\label{sec:lower-bounds}

We first prove logarithmic lower bounds on diamonds, matching the
three approximation guarantees.  We then prove the support tradeoff
and explain why the distance and capacity distributions may differ.
None of these lower bounds uses the weighted upper theorem.

\subsection{Diamond graphs force logarithmic stretch and cut distortion}
\label{subsec:diamond-lower}

The logarithmic order is already necessary on unweighted series-parallel
graphs.  Gupta, Newman, Rabinovich, and Sinclair~\cite{GNRS} proved this
for probabilistic approximation by arbitrary dominating tree metrics,
using recursive diamond graphs.  The argument below is a direct proof
for spanning trees.  Its uniform edge-average form also proves
the corresponding lower bounds for fundamental-cut capacities.

Let $D_0$ consist of one edge with designated terminals.  For $h\geq1$,
form $D_h$ from four disjoint copies of $D_{h-1}$ by putting two copies
in series on each of two parallel branches.  All edges of the resulting
graph have length one and capacity one.  Thus
\[
 m_h=|E(D_h)|=4^h,
 \qquad
 n_h=|V(D_h)|=\frac{2\cdot4^h+4}{3}.
\]
The canonical copies of $D_h$ inside $D_k$ will be called blocks of
height $h$; a block at depth $i$ has height $k-i$.  Only the two terminals
of a block can have incident edges outside that block.

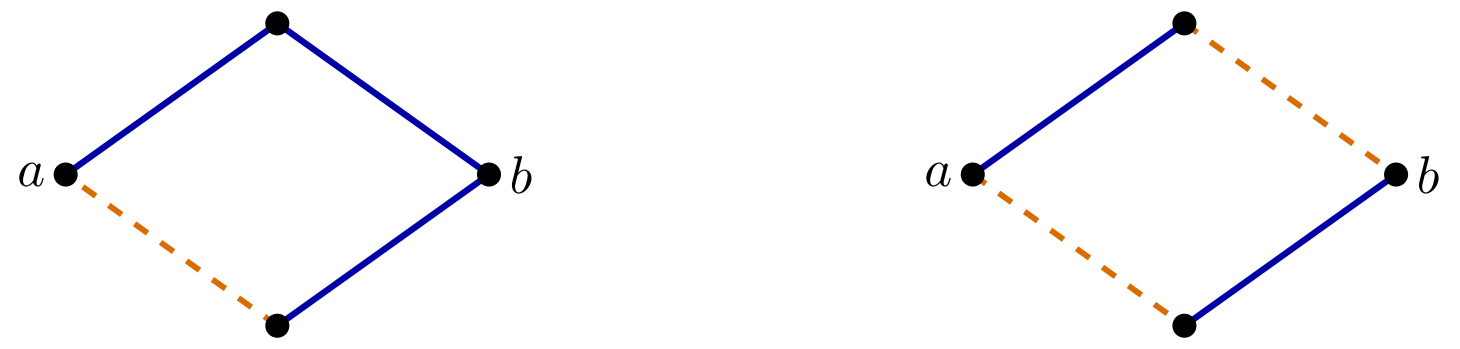
\begin{figure}[htbp]
\centering
\begin{tikzpicture}[x=1cm,y=1cm,
  vertex/.style={circle,fill=black,inner sep=1.6pt},
  every node/.style={font=\small}]
  \begin{scope}
    \coordinate (a) at (0,0);
    \coordinate (t) at (1.4,1);
    \coordinate (b) at (2.8,0);
    \coordinate (d) at (1.4,-1);
    \draw[blue!65!black,line width=1.2pt] (a)--(t)--(b)--(d);
    \draw[orange!85!black,dashed,line width=1.1pt] (d)--(a);
    \foreach \v in {a,t,b,d} {\node[vertex] at (\v) {};}
    \node[left] at (a) {$a$};
    \node[right] at (b) {$b$};
    \node at (1.4,-1.55) {closed parent: one open child};
  \end{scope}
  \begin{scope}[xshift=6cm]
    \coordinate (a) at (0,0);
    \coordinate (t) at (1.4,1);
    \coordinate (b) at (2.8,0);
    \coordinate (d) at (1.4,-1);
    \draw[blue!65!black,line width=1.2pt] (a)--(t) (d)--(b);
    \draw[orange!85!black,dashed,line width=1.1pt] (t)--(b) (a)--(d);
    \foreach \v in {a,t,b,d} {\node[vertex] at (\v) {};}
    \node[left] at (a) {$a$};
    \node[right] at (b) {$b$};
    \node at (1.4,-1.55) {open parent: two open children};
  \end{scope}
\end{tikzpicture}
\caption{The recursive diamond and the component count.  Each side
represents a height-$(h-1)$ child block of a height-$h$ diamond.
A solid side denotes a connected restriction of the spanning tree;
a dashed side denotes a restriction with two components, one at
each child terminal.  A parent with $c$ components has exactly $c$
open children.  The two drawings show the resulting forests on the
four corner vertices.}
\label{fig:diamond-components}
\end{figure}

\begin{proposition}[Uniform average stretch on diamonds]
\label{prop:diamond-average-lower}
For every $k\geq1$ and every spanning tree $T\subseteq D_k$,
\[
 \sum_{e=uv\in E(D_k)}d_T(u,v)
 \geq \frac{k4^k}{6}-\frac{4^k-1}{18}
 \geq \frac{k4^k}{9}.
\]
\end{proposition}

\begin{proof}
We first record two features of a height-$h$ block $B$ with terminals
$a,b$.  Give the two terminals of $D_k$ integer coordinates $0$ and
$2^k$.  Recursively, when the terminals of a height-$h$ block have
coordinates $x$ and $x+2^h$, give both of its intermediate branch
vertices coordinate $x+2^{h-1}$, and continue within the four children.
This defines a single coordinate on every vertex of $D_k$.  Every
original edge changes the coordinate by exactly one.  Consequently
every path in the whole graph between $a$ and $b$ has length at least
$2^h$, even if it leaves $B$.

The block $B$ also contains $2^h$ edge-disjoint paths between its
terminals.  This follows by induction: each pair of series children
supplies $2^{h-1}$ paths, by concatenating their edge-disjoint path
families at the common terminal; the two parallel branches are
edge-disjoint.  Hence every cut inside $B$ separating its terminals
contains at least $2^h$ edges.

Fix $T$, and let $T_B$ be the forest whose vertices are those of $B$
and whose edges are $E(T)\cap E(B)$.  Every component of $T_B$ contains
a terminal of $B$.  Indeed, a component with neither terminal has no
incident edge leaving $B$, and has no tree edge to another component
inside $B$, so it would be disconnected from the rest of $T$.
It follows that $T_B$ has either one or two components.  Call $B$
\emph{closed} in the first case and \emph{open} in the second.  An
open block has one terminal in each of its two components.

Suppose a nonleaf block has $c\in\{1,2\}$ components in its
restricted forest and $o$ open children.  Each closed child contributes
$n_{h-1}-1$ tree edges, and each open child contributes $n_{h-1}-2$.
Since $|V(B)|=4n_{h-1}-4$, counting edges gives
\[
 |E(T_B)|=(4-o)(n_{h-1}-1)+o(n_{h-1}-2)
          =|V(B)|-o.
\]
The same forest has $|V(B)|-c$ edges.  Thus $o=c$: a closed parent
has one open child, and an open parent has two
(Figure~\ref{fig:diamond-components}).

Let $O_i$ denote the number of open blocks at depth $i$.  The root is
closed, and there are $4^i$ blocks at depth $i$.  The preceding count
therefore gives
\[
 O_0=0,
 \qquad
 O_{i+1}=(4^i-O_i)+2O_i=4^i+O_i,
 \qquad
 O_i=\frac{4^i-1}{3}.
\]

For each open block $B$ of height $h$, consider the cut between the
two components of $T_B$ and charge $2^h$ to every edge of $B$ crossing
that cut.  There are at least $2^h$ such edges, so this block creates
total charge at least $4^h$.

If $e=uv$ receives this charge, its endpoints lie in different
components of $T_B$.  The unique $T$-path from $u$ to $v$ must pass
through both terminals of $B$: the only way out of either component
is through its own terminal.  The integer coordinate just constructed
shows that the terminal-to-terminal part of this path has length at
least $2^h$.  In particular, $d_T(u,v)\geq2^h$.

The blocks containing a fixed original edge form a single ancestor
chain, with one block of each height.  The charges received by that
edge thus have distinct values among $1,2,4,\ldots,2^{k-1}$.  Their
sum is less than twice the largest received charge, and that largest
charge is at most $d_T(u,v)$.  (An edge receiving no charge contributes
zero.)  Writing $Q$ for the total charge gives
\[
 2\sum_{e=uv\in E(D_k)}d_T(u,v)
 \geq Q
 \geq\sum_{i=1}^{k}O_i4^{k-i}
 =\frac{k4^k}{3}-\frac{4^k-1}{9}.
\]
Dividing by two proves the first inequality.  The second follows
because their difference is $((k-1)4^k+1)/18\geq0$.
\end{proof}

\begin{corollary}[Three logarithmic lower bounds on the same diamonds]
\label{cor:diamond-three-lower}
Let $\mathcal D$ be any distribution over spanning trees of
$D_k$, with unit original edge lengths and capacities and with the
fundamental-cut capacities defined above.  Then
\begin{align*}
 \max_{uv\in E(D_k)}
 \mathbb E_{T\sim\mathcal D}
   \frac{d_T(u,v)}{d_{D_k}(u,v)}
 &\geq \frac{k}{9},\\
 \max_{f\in E(D_k)}\mathbb E_{T\sim\mathcal D} g_T(f)
 &\geq \frac{k}{9},\\
 \max_{v\in V(D_k)}
 \frac{\mathbb E_{T\sim\mathcal D}
       c_T(\delta_T(\{v\}))}{c_{D_k}(\delta_{D_k}(\{v\}))}
 &\geq \frac{k}{9}.
\end{align*}
In particular, expected stretch, maximum expected congestion, and
maximum expected cut distortion each require order $\log n_k$.
The cut lower bound already holds on singleton cuts.
\end{corollary}

\begin{proof}
For the metric assertion, $d_{D_k}(u,v)=1$ for every original edge.
Average the proposition over $T$, and then average over the $m_k$
original edges.

For capacities, an original edge $uv$ crosses the fundamental cut of
$f\in T$ exactly when the $T$-path from $u$ to $v$ uses $f$.
Double counting gives the pointwise identity
\[
 \sum_{f\in E(D_k)}L_T(f)
 =\sum_{uv\in E(D_k)}|E(P_T(u,v))|
 =\sum_{uv\in E(D_k)}d_T(u,v).
\]
All $c_f=1$.  Averaging over $T$ and then over $f$ proves the congestion
assertion.

Finally, sum the singleton tree cuts.  Every selected tree edge is
counted at its two endpoints, so
\[
 \sum_{v\in V(D_k)}c_T(\delta_T(\{v\}))
 =2\sum_fL_T(f),
 \qquad
 \sum_{v\in V(D_k)}\deg_{D_k}(v)=2m_k.
\]
Therefore the degree-weighted average of the expected singleton-cut
ratios is
\[
 \sum_v\frac{\deg_{D_k}(v)}{2m_k}
 \frac{\mathbb E c_T(\delta_T(\{v\}))}{\deg_{D_k}(v)}
 =\frac{1}{m_k}\mathbb E\sum_fL_T(f)
 \geq\frac{k}{9}.
\]
At least one ratio is at least this average.  This argument does not
require bounded degrees.  Since $n_k=(2\cdot4^k+4)/3$, we have
$k=\Theta(\log n_k)$.
\end{proof}

The witness edge or singleton vertex may depend on the distribution.
The last two assertions use the prescribed fundamental-cut capacities
on spanning trees; they do not assert a lower bound for arbitrary
capacities on arbitrary abstract trees.
\subsection{A sharp support tradeoff for fundamental-cut congestion}
\label{sec:support-size-obstruction}

\begin{figure}[htbp]
\centering
\begin{tikzpicture}[x=1cm,y=0.85cm,
  vertex/.style={circle,fill=black,inner sep=1.8pt},
  every node/.style={font=\small}]
  \coordinate (a) at (1,0);
  \coordinate (au) at (0,1.4);
  \coordinate (ad) at (0,-1.4);
  \coordinate (b) at (6,0);
  \coordinate (bu) at (7,1.4);
  \coordinate (bd) at (7,-1.4);
  \draw[black!25,densely dashed] (au)--(bu) (au)--(b) (au)--(bd)
      (ad)--(bu) (ad)--(b) (ad)--(bd) (a)--(bu) (a)--(bd);
  \draw[blue!65!black,line width=1.5pt] (au)--(a)--(ad)
      (bu)--(b)--(bd);
  \draw[orange!85!black,line width=1.7pt] (a)--(b);
  \foreach \v in {a,au,ad,b,bu,bd} {\node[vertex] at (\v) {};}
  \node[left=4pt] at (a) {$a_0$};
  \node[right=4pt] at (b) {$b_0$};
  \node at (-0.2,0) {$\vdots$};
  \node at (7.2,0) {$\vdots$};
  \node at (0.6,1.95) {$A$: $k$ vertices};
  \node at (6.4,1.95) {$B$: $k$ vertices};
  \node[fill=white,inner sep=2pt] at (3.5,0.35) {selected bridge};
  \node at (3.5,-1.95) {star capacities $k^2$; cross-edge capacities $1$};
\end{tikzpicture}
\caption{The support lower bound.  All $k^2$ cross edges are present;
only representative leaves and cross edges are drawn.  A canonical
tree retains both heavy stars and one bridge.  That bridge carries
the capacity-$k^2$ fundamental cut $A\mid B$.  A mixture spreads this
load among different original cross edges.}
\label{fig:heavy-stars}
\end{figure}
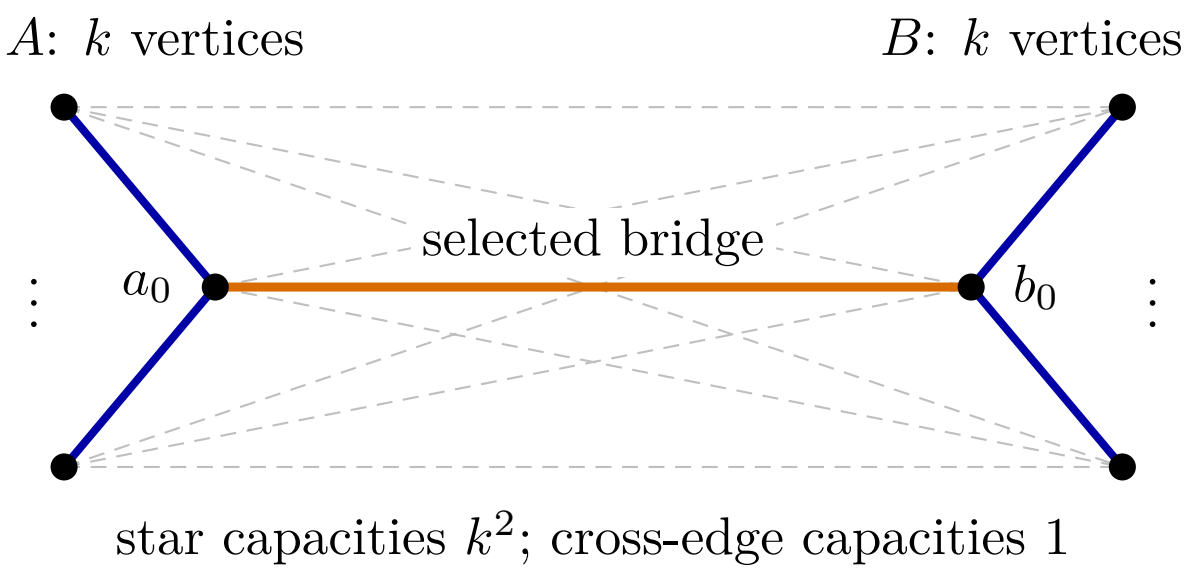

Let $A$ and $B$ be disjoint sets of $k\geq2$ vertices.  Put a star
$H_A$ on $A$ and a star $H_B$ on $B$, assign capacity $M=k^2$ to
every star edge, and add all $k^2$ edges between $A$ and $B$ with
capacity one.  This is the graph $G_k$ of
Theorem~\ref{thm:support-intro}; it has
$n=2k$ vertices and $m=k^2+2k-2$ edges.
For a cross edge $ab$, its \emph{canonical tree} is
\[
 T_{ab}=H_A\cup H_B\cup\{ab\}.
\]
Figure~\ref{fig:heavy-stars} illustrates the construction.

\begin{proof}[Proof of Theorem~\ref{thm:support-intro}]
\emph{Every atom has small probability.}
Every spanning tree contains a cross edge.  Every cut separating
the endpoints of a cross edge either splits a heavy star, and
therefore crosses an edge of capacity $M$, or is the cut $A\mid B$,
whose capacity is also $M$.  Thus each tree $T$ has a cross edge
$f\in T$ with $g_T(f)=L_T(f)/c_f\geq M$.
If the tree has probability $q_T$, nonnegativity gives
\[
 \E g_{T'}(f)\geq q_Tg_T(f)\geq q_TM.
\]
A congestion bound $\rho$ therefore requires $q_T\leq\rho/M$
for every tree, and hence
$|\supp q|\geq\lceil M/\rho\rceil$.

\par\smallskip\noindent\emph{Congestion approaching one.}
Choose a canonical tree $T_{ab}$ by choosing its bridge.
Its bridge has congestion $M$.  Let $f$ be a heavy spoke with
leaf $v\in A$, and let $r_v$ be the probability that the bridge's
$A$ endpoint is $v$.  If the endpoint is elsewhere, the
fundamental-cut capacity of $f$ is $M+k$.  If the endpoint is $v$,
it is $M+k(k-1)$.  Consequently
\begin{equation}
\label{eq:heavy-spoke-congestion}
 \E g_T(f)=1+\frac1k+\frac{k-2}{k}\,r_v.
\end{equation}
The identical formula holds on side $B$ with the column marginal.
The uniform distribution over all $k^2$ bridges has every row and
column marginal equal to $1/k$.  Every cross edge then has expected
congestion one, and every heavy spoke has expected congestion
\[
 \beta_k=1+\frac{2(k-1)}{k^2}.
\]
For any capacitated graph and every spanning tree,
\[
 \sum_f c_fg_T(f)=\sum_e c_e|E(P_T(e))|\geq\sum_e c_e.
\]
Taking expectations shows that maximum expected congestion is
always at least one.  The optimum on $G_k$ is therefore between
$1$ and $\beta_k$, and tends to one.

\par\smallskip\noindent\emph{Attaining the support lower bound.}
Fix $\rho\geq\gamma_k$ and put $r=\lceil k^2/\rho\rceil$.
Choose $r$ distinct cross edges whose degrees on either side are
at most $\lceil r/k\rceil$.  To construct them, decompose
$K_{k,k}$ into its $k$ cyclic perfect matchings, take
$\lfloor r/k\rfloor$ whole matchings, and then take the required
number of edges from the next matching.  Use the corresponding
$r$ canonical trees uniformly.

Every selected cross edge has expected congestion $k^2/r\leq\rho$.
Each row and column marginal is at most
$\lceil r/k\rceil/r\leq1/k+1/r$.  By
\eqref{eq:heavy-spoke-congestion}, every spoke has expected
congestion at most
\[
 1+\frac{2(k-1)}{k^2}+\frac{k-2}{kr}
 \leq1+\frac{2(k-1)}{k^2}+\frac{\rho}{k^2}
 \leq\rho.
\]
The first inequality uses $r\geq k^2/\rho$ and $(k-2)/k\leq1$;
the last is equivalent to $\rho\geq\gamma_k$.
Together with the atom bound, this proves the exact support tradeoff.

\par\smallskip\noindent\emph{Cut and metric approximation with one tree.}
Fix any canonical tree and equip it with fundamental-cut capacities.
Each heavy spoke has tree capacity at most $2M$, and the bridge
has tree capacity $M$.  For a cut $S$, let $h$ be the number of
heavy spokes crossing it.  If $h=0$, the cut is trivial or is
$A\mid B$, and its tree and graph capacities are equal.
If $h\geq1$, then
\[
 c_T(\delta_T(S))\leq2Mh+M\leq3Mh
 \leq3c_G(\delta_G(S)).
\]
The reverse inequality follows from fundamental-cut domination.
Thus one canonical tree gives factor-three approximation of every cut.

For the metric assertion, assign unit length to every edge and choose
the bridge joining the two star centers.  The resulting spanning
tree has diameter three.  Distinct vertices are at graph distance
at least one, so its distance distortion is at most three.
\end{proof}

\begin{corollary}[Nearly $m$ trees for near-unit congestion]
\label{cor:support-near-full}
For the family $G_k$, congestion at most
$\beta_k=1+2(k-1)/k^2$ is attainable, and every distribution with
congestion at most $\beta_k$ has support at least
\[
 \left\lceil\frac{k^4}{m}\right\rceil
 \geq m-2n+4=(1-o(1))m.
\]
The same lower bound applies to an optimal-congestion distribution.
\end{corollary}
\begin{proof}
Here $m=k^2+2k-2$ and $\beta_k=m/k^2$.
Apply Theorem~\ref{thm:support-intro} with $\rho=\beta_k$ and use
\[
 \frac{k^4}{m}
 =m-4k+4+\frac{(2k-2)^2}{m}
 \geq m-2n+4.
\]
Since $m=\Theta(n^2)$, the bound is $(1-o(1))m$.
The optimum congestion is at most $\beta_k$, so every optimal
distribution also satisfies the premise.
\end{proof}

The support lower bound uses the guarantee on each original edge.
It therefore distinguishes that guarantee from expected cut
approximation.  It also separates the two orders of maximum and
expectation: every tree of $G_k$ has congestion at least $k^2$,
whereas one fixed mixture has maximum expected congestion
$1+O(1/k)$.
\subsection{Why a single distribution need not achieve both guarantees}

The preceding results construct a distance distribution for a fixed
length vector and a capacity distribution for a fixed capacity
vector.  One cannot require the same distribution to achieve both
logarithmic guarantees for arbitrary prescribed data.  The next
general family is the obstruction in
Andersen--Feige~\cite[Section 3.4]{AF}.

\begin{proposition}[A simultaneous stretch--congestion obstruction]
\label{prop:cap-simultaneous-obstruction}
For every integer $k\geq2$ there is a graph on
$N=k(k-1)+2$ vertices, with unit lengths and unit capacities, such
that every distribution over its spanning trees has either
maximum expected stretch or maximum expected congestion at least
$(k+1)/2=\Omega(\sqrt N)$.
\end{proposition}

\begin{proof}
Take $k$ internally vertex-disjoint $s$--$t$ paths, each consisting
of $k$ edges, and add the direct edge $e_0=st$.  Write
$q=\Pr[e_0\in T]$.  If $e_0\notin T$, the unique tree path
between $s$ and $t$ is one complete long path and has length $k$;
if $e_0\in T$ it has length one.  Thus
\[
 \mathbb E s_T(e_0)=k-(k-1)q.
\]
If $e_0\in T$, exactly one edge is missing from each long path:
two missing edges on a path would disconnect an internal segment,
and no missing edge would leave a cycle with $e_0$.  Each of those
$k$ missing edges has its tree path traverse $e_0$, in addition
to the path of $e_0$ itself.  Hence $L_T(e_0)=k+1$ in this case.
If $e_0\notin T$, its load as a graph edge is zero.  Therefore
\[
 \mathbb E g_T(e_0)=(k+1)q.
\]
When $q\leq1/2$ the expected stretch is at least $(k+1)/2$;
when $q\geq1/2$ the expected congestion is at least $(k+1)/2$.
This proves the assertion for every $k$ and every distribution.
\end{proof}

\section{Choice of parameters}
\label{sec:choice-of-parameters}

Fix
\[
 \varepsilon=\frac1{100},\qquad
 \delta=\frac1{10000},\qquad
 \chi=8,\qquad A_0=1024,\qquad \zeta=2^{-10}.
\]
We verify the sufficient conditions in
\eqref{eq:admissible-parameters}:
$\varepsilon\leq1/8$, $\delta\leq1/256$, $\chi\geq4$,
$A_0=1024\geq54\chi=432$, and $\zeta=2^{-10}$.
These are the only restrictions on the five parameters used in the
entropy and radius analysis.

For clarity, their roles can also be checked directly.
First, $\varepsilon\leq1/8$ implies
\[
 2\varepsilon<1,\qquad
 \frac{2}{1-2\varepsilon}<3,\qquad
 \frac{1-\varepsilon}{1+\varepsilon}>\frac1\chi.
\]
These inequalities justify geometric summation of nonseed runs, the
uniform source-scale comparison, and the ineligibility of far captures.
Second, $\delta\leq1/256$ gives
\[
 4\delta<\frac12,\qquad
 64\delta\leq\frac14,\qquad
 128\delta\leq\frac12<\frac4{3\varepsilon}.
\]
Thus source cores lie in every allowed cell, low-entropy seeds are
unflagged, and their inherited successor charges contradict the capture
floor.  Third, $\chi>3$ makes the far mass gap
$g_\chi=(\chi-3)/(12\chi)$ positive.
The condition $A_0\geq54\chi$ contains each cell's near set in the
outer ball of any represented near vertex.  It also contains far source
cores, since $3(3+\delta\varepsilon/\chi)<10<A_0$.
Finally $\zeta\leq2^{-10}<\log2$ gives $K_v\leq2$ at a
low-entropy seed, while the complement of its inner ball has mass
strictly below $2^{-10}M$.  Source eligibility and integrality give a
source core of mass at least $2^{-10}M$, so the two cores intersect.

The resulting entropy stride and constants are
\[
 L=1+\left\lceil\log_2\frac{A_0}{\delta\varepsilon}\right\rceil
   =31,\qquad
 C_F=714240<10^6,\qquad C_T=238080<3\cdot10^5.
\]
Indeed $2^{29}<A_0/(\delta\varepsilon)=1024000000<2^{30}$.
Using $\log2<3/4$, we have
\[
 C_R<8,\qquad
 C_S<2400(1+768)<2\cdot10^6,\qquad
 L+1+\frac{\varepsilon C_F}{4\chi}=255.2<256.
\]
Consequently
\[
 C_M<8\bigl(10^6+12\cdot2\cdot10^6\cdot256\bigr)<10^{11}.
\]

The local constants are determined by geometry:
$C_{\rm geo}=34$ and $W=3C_{\rm geo}=102$.
Using $\log2>1/2$ and the preceding bounds in
Theorem~\ref{thm:dyad-weighted} gives
\[
 C_{\rm det}
 =\frac{34(32\chi/\varepsilon+1)}{\log2}
   +3\left[8W(C_M+C_F)+4WC_T
                         +\frac{216\chi W}{\varepsilon}\right]
 <3\cdot10^{14}<10^{15}.
\]
We may therefore take $C_0=10^{15}$ in Theorem~\ref{thm:main}.
The following rational integer-valued response bound is valid:
\[
 \rho_n=10^{15}\left\lceil\log_2(2n)\right\rceil,
 \qquad
 \sum_{e=uv}a_ed_T(u,v)\leq\rho_n\sum_ea_e\ell_e.
\]
The large displayed constant is used only to make the exact rational
master fully explicit; no numerical optimization is needed for the
$O(\log n)$ guarantee.

\paragraph{AI Use Statement.}
This work resulted from extensive interaction between the author and
OpenAI Astra, with the author providing suggestions and critically
checking Astra's responses through iterative exchanges. OpenAI Astra
was also used in writing and revising the manuscript through detailed
prompts from the author, with Gemini Argon assisting in some parts.
The author directed the process and evaluated the proposed ideas and
arguments. A repository link for Lean files formalizing the main
theorems are available on Github at \url{https://github.com/hajiaghayi/Spanning-Tree-Embedding/blob/8dbf5dc3abc88ae1bc7074ff58482eab1aea73e8/SpanningTreeDecompositions-Theorem-1.1.zip}.

\bibliographystyle{alpha}
\bibliography{spanning_tree_decompositions}
\end{document}